\documentclass[journal]{IEEEtran}
\usepackage{stmaryrd}
\usepackage{amsfonts}
\usepackage{amsmath}
\usepackage{amssymb}
\usepackage{cite}
\usepackage{graphicx}
\usepackage{subfigure}
\usepackage{amsthm}
\usepackage{algorithmic}
\usepackage{algorithm}
\usepackage{stfloats}
\usepackage{textcomp}

\newtheorem{Theorem}{Theorem}

\newtheorem{Lemma}[Theorem]{Lemma}
\newtheorem{Example}{Example}

\newcommand{\nchoosek}[2]{ #1 \choose #2}

\newcommand{\abs}[1]{\left|#1\right|}
\newcommand{\parenv}[1]{\left( #1 \right)}
\newcommand{\newQED}{\QED \vspace{0.1in}}

\begin{document}
%
\title{Nonuniform Codes for Correcting Asymmetric Errors in Data Storage}
%
%
%

\author{Hongchao~Zhou,
        Anxiao~(Andrew)~Jiang,~\IEEEmembership{Member,~IEEE,}
        Jehoshua~Bruck,~\IEEEmembership{Fellow,~IEEE}
\thanks{This work was supported in part by the NSF CAREER Award CCF-0747415, the NSF
grant ECCS-0802107, and by an NSF-NRI award. This paper was presented in part at IEEE International Symposium on Information Theory (ISIT), St. Petersburg, Russia, August 2011.}
\thanks{H. Zhou and J. Bruck are with the Department
of Electrical Engineering, California Institute of Technology, Pasadena, CA, 91125.
{\it Email: hzhou@caltech.edu, bruck@caltech.edu}}
\thanks{A. Jiang is with the Computer Science and Engineering Department,
Texas A\&M University,
College Station, TX 77843. {\it Email: ajiang@cse.tamu.edu}}
}

\maketitle

\begin{abstract}
The construction of asymmetric error correcting codes is a topic that was studied extensively, however, the existing approach
for code construction assumes that every codeword should tolerate $t$ asymmetric errors.
Our main observation is that in contrast to symmetric errors, asymmetric errors are content dependent.
For example, in Z-channels, the all-1 codeword is prone to have more errors than the all-0 codeword.
This motivates us to develop nonuniform codes whose codewords can tolerate different
numbers of asymmetric errors depending on their Hamming weights.
The idea in a nonuniform codes' construction is to augment the redundancy in a content-dependent way and
guarantee the worst case reliability while maximizing the code size. In this paper, we first study nonuniform codes for Z-channels, namely, they only suffer one type of errors, say $1\rightarrow 0$. Specifically, we derive their upper bounds, analyze their asymptotic performances, and introduce two general constructions.
Then we extend the concept and results of nonuniform codes to general binary asymmetric channels, where the error probability for each bit from $0$ to $1$ is smaller than that from $1$ to $0$.
\end{abstract}



%
\IEEEpeerreviewmaketitle

\section{Introduction}
%
%
%
%

\IEEEPARstart{A}{symmetric} errors exist in many storage devices \cite{Cassuto2010}. In optical disks, read only memories and quantum memories, the error probability from $1$ to $0$ is significantly higher than the error probability from $0$ to $1$, which is modeled by Z-channels where the transmitted sequences only suffer one type of errors, say $1\rightarrow 0$. In some other devices, like flash memories and phase change memories, although the error probability from $0$ to $1$ is still smaller than that from $1$ to $0$, it is not ignorable. That means both types of errors, say $1\rightarrow 0$ and $0\rightarrow 1$ are possible, modeled by binary asymmetric channels. In contrast to symmetric errors, where the error probability of a codeword is context independent (since the error probability for 1s and 0s is identical), asymmetric errors are context dependent. For example, the all-1 codeword is prone to have more errors than the all-0 codeword in both Z-channels and binary asymmetric channels.

\begin{figure}[!t]
\centering
\includegraphics[width=3in]{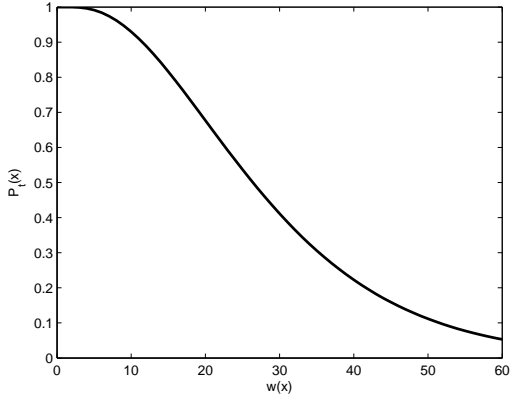}
\caption{The relation between $P_t(\mathbf{x})$ and $w(\mathbf{x})$ when $p=0.1$ and $t=2$.}
\label{fig_errorProbability}
\end{figure}

The construction of asymmetric error correcting
codes is a topic that was studied extensively.
In \cite{Klove1995}, Kl{\o}ve summarized and presented several such codes. In addition, a large amount of efforts are contributed to the design of systematic codes \cite{Abdel-Chaffar1998,Bose2000}, constructing single or multiple error-correcting codes \cite{Al-Bassam1997,Saitoh1990,Tallini08}, increasing
the lower bounds \cite{Etzion1991,Fang1992,Zhang1992, Fu2003} and applying LDPC codes in the context of asymmetric channels \cite{Wang2005}.
However, the existing approach for code construction is similar
to the approach taken in the construction of symmetric error-correcting codes, namely, it assumes that
every codeword could tolerate $t$ asymmetric errors (or generally $t_1$ $1\rightarrow 0$ errors and $t_2$ $0\rightarrow 1$ errors).
As a result, different codewords might have different reliability.
To see this, let's consider errors to be i.i.d., where every bit that is a 1 can change to a 0 by an asymmetric error with crossover probability $p>0$ and each bit that is a $0$ keeps unchanged. For a codeword $\mathbf{x}=\parenv{x_{1},x_{2},\dots,x_{n}} \in \{0,1\}^{n}$, let
$w(\mathbf{x})=\abs{\{i~:1\le i \le n,x_{i}=1\}}$ denote the
Hamming weight of $\mathbf{x}$. Then the probability for
$\mathbf{x}$ to have at most $t$ asymmetric errors is
$$P_t(\mathbf{x})=\sum_{i=0}^{t}
{\nchoosek{w(\mathbf{x})}{i}} p^i (1-p)^{w(\mathbf{x})-i}.$$
Since
$\mathbf{x}$ can correct $t$ errors, $P_t(\mathbf{x})$ is the
probability of correctly decoding $\mathbf{x}$ (assuming codewords
with more than $t$ errors are uncorrectable). It can be
readily observed that the reliability of codewords decreases when their Hamming
weights increase, for example, see Fig. \ref{fig_errorProbability}.

While asymmetric errors are content dependent, in most applications of data storage the reliability of each codeword should be content independent.
Namely, unaware of data importance,  no matter what content is stored, it should be retrieved with very high probability.
The reason is that once a block cannot be correctly decoded, the content of the block, which might be very important, will be lost forever.
So we are interested in the worst-case performance rather than the average performance that is commonly considered in telecommunication, and we want to construct error-correcting codes that can guarantee the reliability of every codeword.
In this case, it is not desired to let all the codewords tolerate the same number of asymmetric errors, since the codeword with the highest Hamming weight will become a `bottleneck' and limit the code rate.
We call the existing codes \emph{uniform codes} while
we focus on the notion of \emph{nonuniform codes}, namely, codes whose
codewords can tolerate different numbers of asymmetric errors
depending on their Hamming weights. The goal of introducing nonuniform
codes is to maximize the code size while guaranteeing the reliability of each codeword for combating asymmetric errors.

In a nonuniform code, given a codeword $\mathbf{x}\in \{0,1\}^n$ of weight $w$, we let $t_{\downarrow}(w)$ denote the number of $1\rightarrow 0$ errors that $\mathbf{x}$ has to tolerate, and we let $t_{\uparrow}(w)$ denote the number of $0\rightarrow 1$ errors that $\mathbf{x}$ has to tolerate. Both $t_{\downarrow}$ and $t_{\uparrow}$ are step functions on $\{0,1,...,n\}$ that can be predetermined by the channel, the types of errors and the required reliability.
In this paper, we consider $t_{\downarrow}$ a nondecreasing function and $t_{\uparrow}$ a nonincreasing function of codeword weight. As a result, we call such a code as a nonuniform code correcting $[t_{\downarrow}, t_{\uparrow}]$ errors. In particular, for Z-channels where $t_{\uparrow}(w)=0$ for all $0\leq w\leq n$, we call it a nonuniform code correcting $t_{\downarrow}$ asymmetric errors. Surprisingly, while nonuniform codes seem to be a natural idea (especially in data storage applications), they were not studied in the literature.

\begin{Example}
In Z-channels, let $p$ be the crossover probability of each bit from $1$ to $0$ and let $q_e<1$ be maximal tolerated error probability for each codeword.
If we consider the errors to be i.i.d., then we can get
\begin{equation}\label{equ_property_1}
t_{\downarrow}(w)=\min\{s\in
N|\sum_{i=0}^s {\nchoosek{w}{i}} p^i(1-p)^{w-i}\geq 1-q_e\}\end{equation}
for $0\leq w\leq n$.
In this case, every erroneous codeword can be corrected with probability at least $1-q_{e}$.\hfill\qed
\end{Example}

The following notations will be  used throughout of this paper:
$$\begin{array}{lcl}
q_e && \textrm{the maximal error probability for each codeword}\\
p,p_{\downarrow} && \textrm{the error probability of each bit from $1$ to $0$}\\
p_{\uparrow} && \textrm{the error probability of each bit from $0$ to $1$}\\
t_{\downarrow} &&\textrm{a nondecreasing function that indicates}\\
&& \textrm{the number of $1\rightarrow 0$ errors to tolerate}\\
t_{\uparrow} &&\textrm{a nonincreasing function that indicates}\\
&& \textrm{the number of $0\rightarrow 1$ errors to tolerate}
\end{array}
$$\vspace{0.02in}

In this paper, we introduce the concept of nonuniform codes and study their basic properties, upper bounds on the rate, asymptotic performance, and code constructions. We first focus on Z-channels and study nonuniform codes correcting $t_{\downarrow}$ asymmetric errors.  The paper is organized as follows: In  Section~\ref{section_properties}, we provide some basic properties of nonuniform codes. In Section~\ref{section_upper},
we give an almost explicit upper bound for the size of nonuniform codes.
Section~\ref{section_asymptotic}
studies and compares the asymptotic performances of nonuniform codes and uniform codes.
Two general constructions, based on multiple layers or bit flips, are proposed in Section~\ref{section_construction1} and Section~\ref{section_construction2}. Finally, we extend our discussions and results from Z-channels to general binary asymmetric channels in
Section~\ref{section_general}, where we study nonuniform codes correcting $[t_{\downarrow},t_{\uparrow}]$ errors, namely, $t_{\downarrow}$ $1\rightarrow 0$ errors and $t_{\uparrow}$ $0\rightarrow1$ errors. Concluding remarks are presented in Section~\ref{section_conclusion}.

\section{Basic Properties of Nonuniform Codes for Z-Channels}
\label{section_properties}

Storage devices such as optical disks, read-only memories and quantum atomic memories can be modeled by Z-channels, in which the information can suffer a single type of error, namely $1\rightarrow 0$.
In this section, we study some properties of nonuniform codes for Z-channels, namely, codes that only correct $t_{\downarrow}$ asymmetric errors.
Typically, $t_{\downarrow}(w)$ is a nondecreasing function in $w$, the weight of the codeword. We prove it in the following lemma
for the case of i.i.d. errors.

\begin{Lemma} \label{lemma_sec1_1}
Assume the errors in a Z-channel are i.i.d., then given any $0<p,q_e<1$, the function $t_{\downarrow}$ defined in (\ref{equ_property_1}) satisfies
 $t_{\downarrow}(w+1)-t_{\downarrow}(w)\in \{0,1\}$ for all $0\leq w\leq n-1$.
\end{Lemma}

\proof Let us define
$$P(k,w,p)=\sum_{i=0}^{k} {\nchoosek{w}{i}}p^i (1-p)^{w-i}.$$
Then
$$
P(k,w,p)=(w-k){\nchoosek{w}{k}}\int_{0}^{1-p} t^{w-k-1}(1-t)^kdt,$$
which leads us to
\begin{eqnarray}
&&P(k,w,p)-P(k,w+1,p)\nonumber \\
&=&\frac{k+1}{w+1}[P(k+1,w+1,p)-P(k,w+1,p)]. \label{equation_1_1}
\end{eqnarray}

First, let us prove that $t_{\downarrow}(w+1)\geq t_{\downarrow}(w)$. Since
$$P(k+1,w+1,p)-P(k,w+1,p)>0,$$
we have $P(k,w,p)>P(k,w+1,p)$.

We know that $P(t_{\downarrow}(w+1),w+1,p)\geq 1-q_e$, so
$$P(t_{\downarrow}(w+1),w,p)>1-q_e.$$
According to definition of $t_{\downarrow}(w)$, we can conclude that
$t_{\downarrow}(w+1)\geq t_{\downarrow}(w)$.

Second, let us prove that $t_{\downarrow}(w+1)-t_{\downarrow}(w)\leq 1$. Based on equation (\ref{equation_1_1}),
we have
\begin{eqnarray*}
&&P(k,w,p)-P(k+1,w+1,p)\\
&=&\frac{w-k}{w+1}[P(k,w+1,p)-P(k+1,w+1,p)].
\end{eqnarray*}
So $P(k,w,p)<P(k+1,w+1,p)$.

We know that $P(t_{\downarrow}(w),w,p)\geq 1-q_e$, therefore
$$P(t_{\downarrow}(w)+1,w+1,p)>1-q_e.$$
According to the definition of $t_{\downarrow}(w+1)$, we have  $t_{\downarrow}(w+1)\leq t_{\downarrow}(w)+1$.

This completes the proof.
\hfill\newQED

Given two binary vectors $\mathbf{x}=(x_{1},\dots,x_{n})$ and
$\mathbf{y}=(y_{1},\dots,y_{n})$, we say $\mathbf{x}\leq
\mathbf{y}$ if and only if $x_i\leq y_i$ for all $1\leq i\leq n$.
Let $\mathcal{B}(\mathbf{x})$ be the (asymmetric) `ball' centered at $\mathbf{x}$, namely,
it consists of all the vectors obtained by changing at most $t_{\downarrow}(w(\mathbf{x}))$
$1$s in $\mathbf{x}$ into $0$s, i.e.,
$$\mathcal{B}(\mathbf{x})=\{\mathbf{v}\in \{0,1\}^n|\mathbf{v}\leq \mathbf{x} \textrm{ and } N(\mathbf{x},\mathbf{v})\leq t_{\downarrow}(w(\mathbf{x}))\},$$
where $w(\mathbf{x})$ is the weight of $\mathbf{x}$ and
$$N(\mathbf{x},\mathbf{y}) \triangleq |\{i:x_i=1,y_i=0\}|.$$

We have the following properties of nonuniform codes as the generalizations of those for uniform codes studied in  \cite{Klove1995}.

\begin{Lemma}~\label{lemma_sec1_2}
Code $C$ is a nonuniform code correcting $t_{\downarrow}$ asymmetric errors
if and only if
$\mathcal{B}(\mathbf{x}) \bigcap
\mathcal{B}(\mathbf{y}) ={\phi}$
for all $\mathbf{x},
\mathbf{y}\in C$ with $\mathbf{x}\neq \mathbf{y}$.
\end{Lemma}

\proof According to the definition of nonuniform
codes, all the vectors in $\mathcal{B}(\mathbf{x})$ can be
decoded as $\mathbf{x}$, and all the vectors in
$\mathcal{B}(\mathbf{y})$ can be decoded as $\mathbf{y}$.
Hence, $\mathcal{B}(\mathbf{x}) \bigcap
\mathcal{B}(\mathbf{y}) ={\phi}$ for all $\mathbf{x},
\mathbf{y}\in C$.
\hfill\newQED

\begin{Lemma}\label{lemma_sec1_4}
There always exists a nonuniform code of the maximum
size that corrects $t_{\downarrow}$ asymmetric errors and
contains the all-zero codeword.
\end{Lemma}

\proof
Let $C$ be a nonuniform code correcting $t_{\downarrow}$ asymmetric errors, and assume that
$00...00 \notin C$.
If there exists a codeword
$\mathbf{x}\in C$ such that $00...00\in
\mathcal{B}(\mathbf{x})$, then we can get a new nonuniform
 code $C'$ of the same size by replacing $\mathbf{x}$
with $00...00$ in $C$. If there does not exist a codeword
$\mathbf{x}\in C$ such that $00...00\in
\mathcal{B}(\mathbf{x})$, then we can get a larger
nonuniform code $C'$ by adding $00...00$ to $C$.
\hfill\newQED

Given a nonuniform code $C$, let $A_r$ denote the number of
codewords with Hamming weight $r$ in $C$, i.e.,
$$A_r=|\{\mathbf{x}\in C| w(\mathbf{x})=r\}|.$$

Given a nondecreasing function $t_{\downarrow}$, let $R_r$ denote a set of weights
that can reach weight $r$ with at most $t_{\downarrow}$ asymmetric errors, namely,
$$R_r=\{0\leq s\leq n| s-t_{\downarrow}(s)\leq r\leq s\}.$$

\begin{Lemma}\label{lemma_sec1_5} Let $C$ be a nonuniform code correcting $t_{\downarrow}$ asymmetric errors. For $0\leq r\leq n$, we have
\begin{equation}\label{equ_property_2}
\sum_{j\in R_r}{\nchoosek{j}{r}}A_{j}\leq {\nchoosek{n}{r}}.
\end{equation}
\end{Lemma}

\proof Let $V_r=\{\mathbf{x}\in \{0,1\}^n| w(\mathbf{x})=r\}$ be the
set consisting of all the vectors of length $n$ and weight $r$.
If $\mathbf{x}\in C$
with $w(\mathbf{x})=j\in R_r$, according to the properties of $t_{\downarrow}$,
$\mathcal{B}(\mathbf{x})$ contains ${\nchoosek{j}{r}}$
vectors of weight $r$, namely
$$|V_r \bigcap \mathcal{B}(\mathbf{x})| = {\nchoosek{j}{r}}.$$
According to Lemma~\ref{lemma_sec1_2}, we know
that $\bigcup_{\mathbf{x}\in C} (V_r \bigcap \mathcal{B}(\mathbf{x}))$ is a disjoint union, in which
the number of vectors is
$$\sum_{j\in R_r}{\nchoosek{j}{r}}A_{j}.$$
Since $\bigcup_{\mathbf{x}\in C} (V_r \bigcap \mathcal{B}(\mathbf{x}))\subseteq V_r$ and there are at most ${\nchoosek{n}{r}}$ vectors in $V_r$, the lemma follows.\hfill\newQED

\section{Upper Bounds}
\label{section_upper}

Let $B_{\alpha}(n,t)$ denote the maximum size of a uniform code correcting $t$ asymmetric errors, and
let $B_{\beta}(n,t_{\downarrow})$ denote the maximum size of a nonuniform code correcting $t_{\downarrow}$ asymmetric errors, where $t$ is a constant and $t_{\downarrow}$ is a nondecreasing function of codeword weight.
In this section, we first present some existing results on the
upper bounds of $B_{\alpha}(n,t)$ for uniform codes. Then we
derive an almost explicit upper bound of $B_{\beta}(n,t_{\downarrow})$ for
nonuniform codes.

\subsection{Upper Bounds for Uniform Codes}

An explicit upper bound to $B_{\alpha}(n,t)$ was given by Varshamov \cite{Varshamov1964}. In \cite{Klove1995},
Borden showed that $B_{\alpha}(n,t)$ is upper bounded by $$\min\{A(n+t,
2t+1), (t+1)A(n, 2t+1)\},$$ where $A(n,d)$ is the
maximal number of vectors in $\{0,1\}^n$ with Hamming distance at
least $d$. Goldbaum \cite{Goldbaum1971} pointed out that the upper bounds can be
obtained using integer programming. By adding more constrains to
the integer programming, the upper bounds were later improved by
Delsarte and Piret \cite{Delsarte1981} and Weber et al.
\cite{Weber1987}\cite{Weber1988}. Kl{\o}ve generalized the bounds
of Delsarte and Piret, and gave an almost explicit upper bound
which is very easy to compute by relaxing some of the
constrains\cite{Klove1981}, in the following way.

\begin{Theorem}\emph{\cite{Klove1981}} For $n>2t\geq 2$, let $y_0, y_1, ..., y_n$ be defined by
\begin{align*}
  1) & \hspace{0.05in}y_0=1, \\
  2) & \hspace{0.05in}y_r=0, \quad \forall 1\leq r\leq t, \\
  3) & \hspace{0.05in}y_{t+r} =\frac{1}{{\nchoosek{t+r}{t}}}[{\nchoosek{n}{r}}-\sum_{j=0}^{t-1}y_{r+j}{\nchoosek{r+j}{ j}}], \forall 1\leq r\leq \frac{n}{2}-t,\\
  4) & \hspace{0.05in}y_{n-r} =y_r, \quad \forall 0\leq r< \frac{n}{2}.
\end{align*}
%
Then $B_{\alpha}(n,t)\leq M_\alpha(n,t)\triangleq \sum_{r=0}^n
y_r$.\label{theorem_sec2_1}
\end{Theorem}

This method obtains a good upper bound to $B_{\alpha}(n,t)$ (although
it is not the best known one). Since it is easy to compute,
when $n$ and $t$ are large, it is every useful for analyzing the
sizes of uniform codes.

\subsection{Upper Bounds for Nonuniform Codes}

We now derive an almost explicit upper bound for the size
of nonuniform codes correcting $t_{\downarrow}$ asymmetric errors, followed the idea of Kl{\o}ve \cite{Klove1981} for uniform codes.
According to the lemmas in the previous section, we can get an upper bound of $B_{\beta}(n,t_{\downarrow})$, denoted by $M_{\beta}(n,t_{\downarrow})$, such that
$$M_\beta(n,t_{\downarrow})=\max \sum_{i=0}^n z_r,$$
where the
maximum is taken over the following constraints:
\begin{align*}
    1)&\hspace{0.05in} z_r \textrm{ are nonnegative real numbers},\quad\quad\quad\quad\quad\quad\quad\quad\quad\\
    2)&\hspace{0.05in} z_0=1,\\
    3)&\hspace{0.05in} \sum_{j\in R_r}{j\choose r}z_{j}\leq {n \choose r} ,\forall 0\leq  r\leq n.
\end{align*}
Here, condition $2)$ is given by Lemma~\ref{lemma_sec1_4}, and
condition $3)$ is given by Lemma~\ref{lemma_sec1_5}. Our goal is to find an almost explicit way to calculate
$M_\beta(n,t_{\downarrow})$.

\begin{Lemma} \label{Lemma_sec2_1} Assume $\sum_{r=0}^n z_r$ is maximized over $z_0, z_1, ..., z_n$ in the problem above.
If $r=s-t_{\downarrow}(s)$ for some integer $s$ with $0\leq s,r\leq n$, then
$$Z_r=\sum_{j\in R_r} {\nchoosek{j}{r}} z_j ={\nchoosek{n}{r}}.$$
\end{Lemma}

\begin{figure}[!t]
\centering
\includegraphics[width=2in]{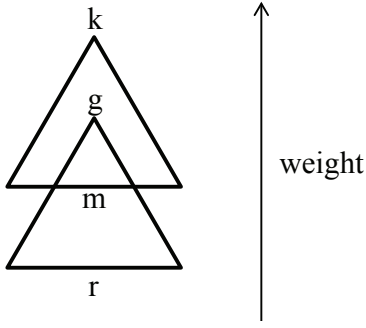}
\caption{This diagram demonstrates the relative values of $r,g,k,m$.} \label{fig_upper}
\end{figure}

\proof Suppose that $Z_r<{\nchoosek{n}{r}}$ for some $r$ that satisfies the above condition. Let $g=\max R_r$ and
$k=\min\{w|z_w>0, w>g\}$, as indicated in Figure \ref{fig_upper}, where a triangular denote the ball centered at the top vertex. Furthermore, we let $m=\max\{w|k-t_{\downarrow}(k)>w\}$.
Note that in this case $r=g-t_{\downarrow}(g)$ and $m=k-t_{\downarrow}(k)-1$.

We first prove that for all
$r\leq w\leq m$, $Z_w<{\nchoosek{n}{w}}$. In order to prove this, we let
$s=w-r$, then we get
\begin{eqnarray*}
  Z_w &=&\sum_{j\in R_w} z_{j}{\nchoosek{j}{w}}\\
   &=&\sum_{j=w}^{g} z_{j}{\nchoosek{j}{w}}\\
  &=& \sum_{j=s}^{g-r} z_{r+j}{\nchoosek{r+j}{r+s}}.
\end{eqnarray*}

It is easy to obtain that $$
{\nchoosek{r+j}{r+s}}={\nchoosek{r+j}{r}} \frac{{\nchoosek{j}{s}}}{{\nchoosek{r+s}{s}}}.$$

So
\begin{eqnarray*}
  Z_w
  &\leq & \frac{{\nchoosek{g-r}{s}}}{{\nchoosek{r+s}{s}}}\sum_{j=s}^{g-r} z_{r+j}{\nchoosek{r+j}{r}}\\
  &<&  \frac{{\nchoosek{g-r}{s}}}{{\nchoosek{r+s}{s}}} {\nchoosek{n}{r}}\\
  &=& \frac{(g-r)(g-r-1)...(g-r-s+1)}{(n-r)(n-r-1)...(n-r-s+1)} {\nchoosek{n}{r+s}}\\
  &\leq& {\nchoosek{n}{w}}.
\end{eqnarray*}

Now, we construct a new group of real numbers $z_0^*, z_1^*, ...,
z_n^*$ such that
\begin{align*}
    1)&\hspace{0.05in}z_g^*=z_g+\Delta,\quad\quad\quad\quad\quad\quad\quad\quad\quad\quad\quad\quad\quad\quad\quad\quad\quad\\
    2)&\hspace{0.05in}z_k^*=z_k-\delta,\\
    3)&\hspace{0.05in}z_r^*=z_r \textrm{ for }r\neq h, r\neq k.
\end{align*}
with
$$ \Delta=\min(\{\frac{{\nchoosek{n}{w}}-Z_w}{{\nchoosek{g}{w}}}|r\leq w\leq m\}\bigcup \{\frac{{\nchoosek{k}{w}}}{{\nchoosek{g}{w}}}z_k|m<w\leq g\}),$$
$$\delta=\frac{1}{\min\{\frac{{\nchoosek{k}{w}}}{{\nchoosek{g}{w}}}|m<w\leq g\}}\Delta.$$

For such $\Delta, \delta$, it is not hard to prove that
$Z_r^*={\nchoosek{n}{r}}$ for $0\leq r\leq n$. On the other hand,
$$\sum_{r=0}^n z_r^* =\sum_{r=0}^n z_r+\Delta-\delta>\sum_{r=0}^n
z_r,$$ which contradicts our assumption that $\sum_{r=0}^n z_r$ is
maximized over the constrains. So the lemma is true. \hfill\newQED \vspace{-0.1in}

\begin{Lemma}\label{lemma_sec2_2}
Assume $\sum_{r=0}^n z_r$ is maximized over $z_0, z_1, ..., z_n$ in the problem above.
If $r=s-t_{\downarrow}(s)$ for some integer $s$ with $0\leq s,r\leq n$, then
$$Z_r=\sum_{j=r}^{h} {\nchoosek{j}{r}} z_j ={\nchoosek{n}{r}},$$
where $$h=\min\{s\in N|s-t_{\downarrow}(s)=r\}.$$
\end{Lemma}

\emph{Sketch of Proof:}  Let
$g=\max\{s\in N|s-t_{\downarrow}(s)=r\}$. If $g=h$,
then the lemma is true. So we only need to prove it for the case
that $g>k$. Similar to
lemma~\ref{Lemma_sec2_1}, we assume $Z_r<{\nchoosek{n}{r}}$, to get the contradiction, we can construct a
new group of real numbers $z_0^*, z_1^*, ..., z_n^*$ such that
\begin{align*}
    1)&\hspace{0.05in}z_{h}^*=z_{h}+\Delta,\quad\quad\quad\quad\quad\quad\quad\quad\quad\quad\quad\quad\quad\quad\quad\quad\quad\\
    2)&\hspace{0.05in}z_w^*=0 \textrm{ for } h<w\leq g,\\
    3)&\hspace{0.05in}z_w^*=z_w \textrm{ if } w\notin [h,g].
\end{align*}
with
$$\Delta=\min\{\frac{\sum_{j=h+1}^{g}
{\nchoosek{j}{w}} z_j}{{\nchoosek{h}{  w}}}| r\leq w\leq
h\}.$$

For this $z_0^*, z_1^*, ..., z_n^*$, they satisfy all the constrains and
$$Z_r^*=\sum_{j=r}^{h} {\nchoosek{j}{r}} z_j ^*={\nchoosek{n}{ r}}.$$ At the same time, it can be proved that
$$\sum_{r=0}^n z_r^*> \sum_{r=0}^n z_r,$$
which contradicts with our assumption that $\sum_{r=0}^n z_r$ is maximized over the constrains.
This completes the proof.
\hfill\newQED

Now let $y_0, y_1, ..., y_n$ be a group of optimal solutions to
$z_0,z_1,...,z_n$ that maximize $\sum_{r=0}^n
z_r$. Then $y_0, y_1, ..., y_n$ satisfy the condition in
Lemma~\ref{lemma_sec2_2}. We see that $y_0=1$. Then based on Lemma~\ref{lemma_sec2_2},
we can get $y_1, ..., y_n$ uniquely by iteration. Hence, we have the following
theorem for calculating the upper bound $M_\beta(n,t_{\downarrow})$.

\begin{Theorem}
Let $y_0, y_1, ..., y_n$ be defined by
\begin{align*}
    1)&\hspace{0.05in}y_0=1,\quad\quad\quad\quad\quad\quad\quad\quad\quad\quad\quad\quad\quad\quad\quad\quad\quad\\
    2)&\hspace{0.05in}y_r=\frac{1}{{r\choose t_{\downarrow}(r)}}[{n \choose r-t_{\downarrow}(r)}-\sum_{j=1}^{t_{\downarrow}(r)}y_{r-j}{r-j \choose t_{\downarrow}(r)-j}],\\
     &\hspace{0.05in}\forall 1\leq r\leq n.
\end{align*}
Then $B_{\beta}(n,t_{\downarrow})\leq M_{\beta}(n,t_{\downarrow})=\sum_{r=0}^n y_r$.\label{theorem_sec2_2}
\end{Theorem}

This theorem provides an almost explicit expression for the upper
bound $M_{\beta}(n,t_{\downarrow})$, which is much easier to calculate than
the equivalent expression defined at the beginning of this
subsection. Note that in the theorem, we do not have a constrain
like the one (constraint 4)  in Theorem~\ref{theorem_sec2_1}. It is
because that the optimal nonuniform codes do not have symmetric weight
distributions due to the fact that $t_{\downarrow}(w)$ monotonically
increases with $w$.

\subsection{Comparison of Upper Bounds}

Here we focus on i.i.d. errors, i.e., given the crossover probability $p$ from $0$ to $1$ and the maximal tolerated error probability $q_e$,
the function $t_{\downarrow}$ is defined in equation (\ref{equ_property_1}). In this case,
we can write the maximum size of a uniform code as $B_{\alpha}(n,t_{\downarrow}(n))=B_{\alpha}(n,p,q_e)$, and
write the maximum size of a nonuniform code as $B_{\beta}(n,t_{\downarrow}(n))=B_{\beta}(n,p,q_e)$.

\begin{figure}[!t]
\centering
\includegraphics[width=3.6in]{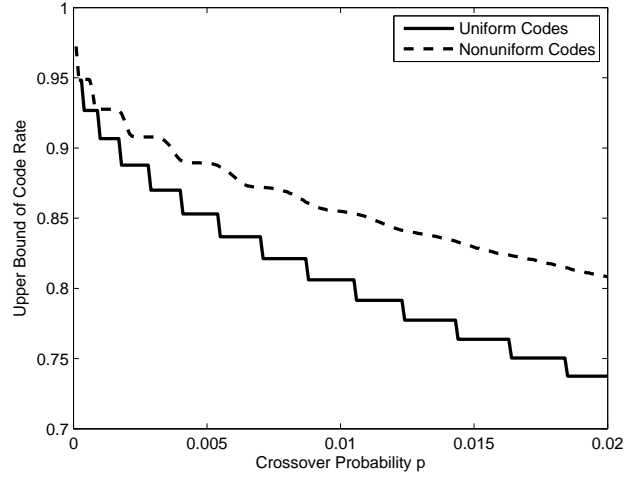}
\caption{Upper bounds of the rates for uniform/nonuniform codes when $n=255, q_e=10^{-4}$.}
\label{fig_upbound2}
\end{figure}

Now we let $\eta_{\alpha}(n,p,q_e)$ denote the maximal code rate defined by
$$\eta_{\alpha}(n,p,q_e)=\frac{\log B_{\alpha}(n,p,q_e)}{n}.$$
Similar, we let $\eta_{\beta}(n,p,q_e)$ denote the maximal code rate defined by
$$\eta_{\beta}(n,p,q_e)=\frac{\log B_{\beta}(n,p,q_e)}{n}.$$
By the definition of uniform and nonuniform codes, it is simple
to see that $\eta_\beta(n,p,q_e) \ge \eta_\alpha(n,p,q_e)$.

Figure \ref{fig_upbound2} depicts the upper bounds of $\eta_{\alpha}(n,p,q_e)$ and $\eta_{\beta}(n,p,q_e)$
for different values of $p$ when $n=255$ and $q_e=10^{-4}$. The upper bound of $\eta_{\alpha}(n,p,q_e)$
is obtained based on the almost explicit upper bound given by Kl{\o}ve, and the upper bound of $\eta_{\beta}(n,p,q_e)$
is obtained based on the almost explicit method proposed in this section.
 It demonstrates that given the same parameters,
the upper bound for nonuniform codes is substantially greater than
that for uniform codes.

\section{Asymptotic Performance}
\label{section_asymptotic}

In this section, we study and compare the asymptotic rates of uniform codes and nonuniform codes.
Note that the performance of nonuniform codes strongly depends on the selection of the function $t_{\downarrow}$.
Here, we focus on i.i.d. errors, so
given $0<p,q_e<1$, we study the asymptotic
behavior of $\eta_\alpha(n,p,q_e)$ and $\eta_\beta(n,p,q_e)$ as
$n\rightarrow \infty$. By the
definition of nonuniform and uniform codes, the `balls' containing
up to $t_{\downarrow}(\mathbf{x})$ (or $t_{\downarrow}(n)$) errors that are centered at
codewords $\mathbf{x}$ need to be disjoint.

Before giving the asymptotic rates, we first present
the following known result:  For any $\delta>0$, when $n$ is large enough, we have
$$2^{n(H(\frac{k}{n})-\delta)}\leq {\nchoosek{n}{ k}}\leq 2^{n(H(\frac{k}{n})+\delta)},$$
where $H(p)$ is the entropy function with
$$H(p)=p\log\frac{1}{p}+(1-p)\log\frac{1}{1-p}\textrm{ for } 0\leq p\leq 1,$$
and
$$H(p)=0 \textrm{ for } p>1 \textrm{ or } p<0.$$

\begin{Lemma}
Let $A(n,d,w)$ be the maximum size of a constant-weight binary
code of codeword length $n$, whose Hamming weight is $w$
and minimum distance is $d$. Let $R(n,t,w)$ be the
maximum size of a binary code with  Hamming weight $w$ and
codeword length $n$ where every codeword can correct $t$
asymmetric errors. Then
$$R(n,t,w)=A(n,2(t+1),w).$$
\end{Lemma}
\proof  Let $C$ be a code of length $n$, constant weight $w$ and
size $R(n,t,w)$ that corrects $t$ asymmetric errors. For all $\mathbf{x}\in \{0,1\}^n$, let's
define $S_t(\mathbf{x})$ be the set consisting of all the vectors
obtained by changing at most $t$ $1$s in $\mathbf{x}$ into $0$s, i.e.,
$$S_t(\mathbf{x})=\{\mathbf{v}\in \{0,1\}^n |\mathbf{v}<\mathbf{x}\textrm{ and } N(\mathbf{x},\mathbf{v})\leq t\}.$$
Then
$\forall
\mathbf{x},\mathbf{y}\in C$, we know that
$S_t(\mathbf{x})\bigcap S_t(\mathbf{y})={\phi}$.

Let $\mathbf{u}=(u_{1},\dots,u_{n})$ be a vector such that
$u_i=\min\{x_i,y_i\}$ for $1\leq i\leq n$. Then
$N(\mathbf{x},\mathbf{u})=N(\mathbf{y},\mathbf{u})$ and
$\mathbf{u}\notin S_t(\mathbf{x})\bigcap S_t(\mathbf{y})$.
W.l.o.g, suppose that $\mathbf{u}\notin S_t(\mathbf{x})$. Then
$N(\mathbf{x},\mathbf{u})>t$, and the Hamming distance between
$\mathbf{x}$ and $\mathbf{y}$ is
$$d(\mathbf{x},\mathbf{y})=N(\mathbf{x},\mathbf{u})+N(\mathbf{y},\mathbf{u})\geq 2(t+1).$$
So the minimum distance of $C$ is at least $2(t+1)$. As a result,
$A(n,2(t+1),w)\geq R(n,t,w)$.

On the other hand, if a constant-weight code has minimum distance
at least $2(t+1)$, it can correct $t$ asymmetric errors. As a
result, $R(n,t,w)\geq A(n,2(t+1),w)$. \hfill\newQED\vspace{-0.1in}

\subsection{Bounds of $\lim_{n\rightarrow\infty}\eta_\alpha(n,p,q_e)$}

Let us first give the lower bound of $\lim_{n\rightarrow\infty}\eta_\alpha(n,p,q_e)$ and then provide
the upper bound.

\begin{Theorem}[Lower bound] \label{theorem_asym1} Given $0<q_e<1$, if $0<p\leq\frac{1}{4}$, we have
$$\lim_{n\rightarrow\infty}\eta_\alpha(n,p,q_e)\geq 1-H(2p).$$
\end{Theorem}

\proof We consider uniform codes that correct $t$ asymmetric errors, where
$$t=\min\{s|\sum_{i=0}^s {\nchoosek{n}{ i}} p^i (1-p)^{n-i}\geq 1-q_e\}.$$

According to Hoeffding's inequality, for any $\delta>0$, as $n$ becomes large enough,
we have $(p-\delta)n\leq t\leq (p+\delta)n$. If we write $t=\gamma n$, then
$p-\delta\leq \gamma\leq p+\delta$ for $n$ large enough.

Since each codeword tolerates $t$ asymmetric errors, we have
$$B_\alpha(n,p,q_e)=B_\alpha(n,t)\geq R(n,t,w)=A(n,2(t+1),w),$$ for
every $w$ with $0\leq w\leq n$. The Gilbert Bound gives that (see
Graham and Sloane\cite{Graham1980})
$$A(n,2(t+1),w)\geq \frac{{\nchoosek{n}{ w}}}{\sum_{i=0}^t {\nchoosek{w}{ i}}{\nchoosek{n-w}{ i}}}.$$

Hence
\begin{eqnarray*}
  B_\alpha(n,p,q_e)& \geq &\max_{w=0}^n \frac{{\nchoosek{n}{w}}}{\sum_{i=0}^t {\nchoosek{w}{i}}{\nchoosek{n-w}{i}}}\\
  &\geq& \max_{w=0}^n\frac{{\nchoosek{n}{w}}}{n\max_{i\in [0,t]} {\nchoosek{w}{i}}{\nchoosek{n-w}{i}}}\\
  &\geq& \max_{w: \frac{w(n-w)}{n}>t }\frac{{\nchoosek{n}{w}}}{n\max_{i\in [0,t]} {\nchoosek{w}{i}}{\nchoosek{n-w}{i}}}\\
  &\geq & \max_{w: \frac{w(n-w)}{n}>t }\frac{{\nchoosek{n}{w}}}{n {\nchoosek{w}{t}}{\nchoosek{n-w}{t}}}.
\end{eqnarray*}

For a binomial term ${\nchoosek{n}{k}}=\frac{n!}{k!(n-k)!}$ and $\delta>0$, when $n$ is large enough,
$$2^{n(H(\frac{k}{n})-\delta)}\leq {\nchoosek{n}{k}}\leq 2^{n(H(\frac{k}{n})+\delta)}.$$

Let $w=\theta n$ and $t=\gamma n$ with $0\leq \theta,\gamma\leq 1$, as $n$ becomes large enough, we have
\begin{eqnarray*}
&&\eta_\alpha(n,p,q_e)\\
&=& \frac{1}{n}\log_2  B_\alpha(n,p,q_e)\\
&\geq &  \frac{1}{n}\log_2  \max_{w: \frac{w(n-w)}{n}>t }\frac{{\nchoosek{n}{w}}}{n {\nchoosek{w}{t}}{\nchoosek{n-w}{t}}}\\
&\geq &  \frac{1}{n}\log_2 \max_{\theta: \theta(1-\theta)>\gamma}\frac{2^{(H(\theta)-\delta)n}}{n 2^{(H(\frac{\gamma}{\theta})+\delta)\theta n}2^{(H(\frac{\gamma}{1-\theta})+\delta)(1-\theta)n}}\\
&\geq & \max_{\theta: \theta(1-\theta)\geq \gamma}H(\theta) - \theta H(\frac{\gamma}{\theta})-(1-\theta)H(\frac{\gamma}{1-\theta}) -2\delta \\
&& \hspace{0.6in}+\frac{1}{n}\log\frac{1}{n}.
\end{eqnarray*}

From $\theta(1-\theta)\geq \gamma$, we get
$\theta>\gamma>0$; then $H(\frac{\gamma}{\theta})$ is a continuous
function of $\gamma$. As $n$ becomes large, we have $p-\delta\leq
\gamma\leq p+\delta$, so we can approximate
$H(\frac{\gamma}{\theta})$ with $H(\frac{p}{\theta})$. Similarly,
we can approximate $H(\frac{\gamma}{1-\theta})$ with
$H(\frac{p}{1-\theta})$. Then we can get as $n\rightarrow\infty$,
\begin{eqnarray*}
&&\eta_\alpha(n,p,q_e)\\
&\geq& \max_{\theta:
\theta(1-\theta)>p}H(\theta) - \theta
H(\frac{p}{\theta})-(1-\theta)H(\frac{p}{1-\theta}).
\end{eqnarray*}

If $0\leq p\leq \frac{1}{4}$, the maximum value can be achieve at
$\theta^*=\frac{1}{2}$. Hence we have
$$\lim_{n\rightarrow\infty}\eta_\alpha(n,p,q_e)\geq 1-H(2p).$$

This completes the proof. \hfill\newQED\vspace{-0.1in}

\begin{Theorem}[Upper bound] \label{theorem_asym2} Given $0<p, q_e<1$, we have
$$\lim_{n\rightarrow \infty}\eta_\alpha(n,p,q_e) \leq (1+p)[1-H(\frac{p}{1+p})].$$
\end{Theorem}

\proof For a uniform code correcting $t$ asymmetric
errors, we have the following observations:
\begin{enumerate}
  \item There is at most one codeword with Hamming weight at most
  $t$;
  \item For $t+1\leq w\leq n$, the number of codewords with Hamming weight $w$ is at most $\frac{{\nchoosek{n}{w-t}}}{{\nchoosek{w}{t}}}$.
\end{enumerate}

Consequently, the total number of codewords is
\begin{eqnarray*}
 B_\alpha(n,p,q_e)&\leq& 1+\sum_{w=t+1}^{n}\frac{{\nchoosek{n}{w-t}}}{{\nchoosek{w}{t}}}  \\
 &=& 1+\sum_{w=t+1}^n \frac{{\nchoosek{n+t}{w}}}{{\nchoosek{n+t}{t}}}\\
 & \leq &  \frac{2^{n+t}}{{\nchoosek{n+t}
 {t}}}.
\end{eqnarray*}

So as $n\rightarrow\infty$, we have
\begin{eqnarray*}
\eta_\alpha(n,p,q_e)&\leq &  \frac{1}{n}\log [\frac{2^{n+t}}{{\nchoosek{n+t}{t}}}]\\
& \leq &  \frac{1}{n}\log \frac{2^{(1+\gamma)n}}{2^{H(\frac{\gamma}{1+\gamma})(1+\gamma)n}}\\
&=& (1+\gamma)-H(\frac{\gamma}{1+\gamma})(1+\gamma)\\
&=& (1+p)[1-H(\frac{p}{1+p})],
\end{eqnarray*}
where the last step is due to the continuousness of $(1+\gamma)-H(\frac{\gamma}{1+\gamma})(1+\gamma)$ over $\gamma$.

This completes the proof.
\hfill\newQED

We see that when $n\rightarrow \infty$, $\eta_\alpha(n,p,q_e)$ does not depends on $q_e$ as long as $0<q_e<1$.
It is because that when $n\rightarrow\infty$, we have $t\rightarrow pn$, which does not depend on $q_e$.
This property is also hold by $\eta_\beta(n,p,q_e)$ when $n\rightarrow \infty$.

\subsection{Bounds of $\lim_{n\rightarrow\infty}\eta_\beta(n,p,q_e)$}

In this subsection, we study the bounds of the asymptotic rates of nonuniform codes. Here, we use the same
idea as that for uniform codes, besides that we need also prove that the `edge effect' can be ignored, i.e., the number of codewords with Hamming weight $w\ll
n$ does not dominate the final result.

\begin{Theorem}[Lower bound]\label{theorem_asym3}
Given $0<p, q_e<1$, we have
$$\lim_{n\rightarrow \infty}\eta_\beta(n,p,q_e)\geq \max_{0\leq \theta \leq 1-p}H(\theta) - \theta H(p)-(1-\theta)H(\frac{p\theta}{1-\theta}).$$
\end{Theorem}

\proof We consider nonuniform codes that corrects $t_{\downarrow}$ asymmetric errors, where
$$t_{\downarrow}(w)=\min\{s|\sum_{i=0}^s {\nchoosek{w}{i}} p^i(1-p)^{w-i}\geq 1-q_e\},$$
for all $0\leq w\leq n$.

Based on Hoeffding's inequality, for any $\delta>0$, as $w$ becomes large enough,
we have $(p-\delta)w\leq t_{\downarrow}(w)\leq (p+\delta)w$. In another word,  for any $\epsilon, \delta>0$,
when $n$ is large enough and $w\geq \epsilon n$, we have $(p-\delta)w\leq t_{\downarrow}(w)\leq (p+\delta)w$.

Let $w=\theta n$ and $t_{\downarrow}(w)=\gamma w$, then when $n$ is large enough, if $\theta>\epsilon$,
we have $$(p-\delta)\leq \gamma \leq (p+\delta).$$
If $\theta<\epsilon$, we call it the `edge' effect. In this case $0\leq \gamma \leq 1$.

Since each codeword with Hamming weight $w$ can tolerate $t_{\downarrow}(w)$ errors, $$B_\beta(n,p,q_e)\geq R(n,t_{\downarrow}(w),w)\geq A(n,2(t_{\downarrow}(w)+1),w),$$
for every $w$ with $0\leq w\leq n$.

Applying the Gilbert Bound, we have
\begin{eqnarray*}
B_\beta(n,p,q_e)&\geq & \max_{w} \frac{{\nchoosek{n}{w}}}{\sum_{i=0}^{t_{\downarrow}(w)} {\nchoosek{w}{i}}{\nchoosek{n-w}{i}}}\\
\end{eqnarray*}

Then
\begin{eqnarray*}
B_\beta(n,p,q_e)&\geq & \max_{w} \frac{{\nchoosek{n}{w}}}{\max_{i\in[0,t_{\downarrow}(w)]}n {\nchoosek{w}{i}}{\nchoosek{n-w}{i}}}\\
  &\geq& \max_{w: \frac{w(n-w)}{n}\geq t_{\downarrow}(w) }\frac{{\nchoosek{n}{w}}}{n {\nchoosek{w}{t_{\downarrow}(w)}}{\nchoosek{n-w}{ t_{\downarrow}(w)}}}.
\end{eqnarray*}

When $n\rightarrow \infty$, we have
\begin{eqnarray*}
&&\eta_\beta(n,p,q_e)\\
&=& \frac{1}{n}\log_2 B_\beta(n,p,q_e)\\
&\geq&  \frac{1}{n}\log_2 \max_{\theta: (1-\theta)\geq \gamma }\frac{2^{(H(\theta)-\delta)n}}{n 2^{(H(\gamma)+\delta)\theta n}2^{(H(\frac{\gamma\theta}{1-\theta})+\delta)(1-\theta)n}}\\
&\geq & \max_{\theta: (1-\theta)\geq \gamma} H(\theta)-\theta H(\gamma)-(1-\theta)H(\frac{\gamma\theta}{1-\theta})\\
&& \hspace{0.6in}-2\delta+\frac{1}{n}\log\frac{1}{n} \\
&= & \max_{\theta: (1-\theta)\geq \gamma} H(\theta)-\theta H(\gamma)-(1-\theta)H(\frac{\gamma\theta}{1-\theta}).
\end{eqnarray*}

Note that when $\theta<\epsilon$ for small $\epsilon$, we have
$$H(\theta)-\theta H(\gamma)-(1-\theta)H(\frac{\gamma\theta}{1-\theta})\sim 0.$$
So we can ignore this edge effect. That implies that we can write
$$p-\delta\leq \gamma \leq p+\delta,$$
for any $\theta$ with $0\leq \theta\leq 1$.

Since $1-\theta\geq \gamma>0$, for any fixed $\theta$, $$H(\theta)-\theta H(\gamma)-(1-\theta)H(\frac{\gamma\theta}{1-\theta})$$ is
a continuous function of $\gamma$. As $n\rightarrow\infty$, we have
$$\eta_\beta(n,p,q_e)\geq \max_{\theta: (1-\theta)\geq p} H(\theta)-\theta H(p)-(1-\theta)H(\frac{p\theta}{1-\theta}).$$

This completes the proof.
\hfill\newQED

\begin{table*}
  \centering
  \caption{Upper bounds and lower bounds for the maximum rates of uniform codes and nonuniform codes}
   \renewcommand{\arraystretch}{2}
  \begin{tabular}{|l|l|l|}
    \hline \rule{0pt}{2ex}
     & Lower Bound & Upper Bound \\
    \hline \rule{0pt}{3ex}
    $\lim_{n\rightarrow \infty}\eta_\alpha(n,p,q_e)$ & $[1-H(2p)]I_{0\leq p\leq \frac{1}{4}}$ &  $ (1+p)[1-H(\frac{p}{1+p})]$\\
   \hline \rule{0pt}{3ex}
   $\lim_{n\rightarrow \infty}\eta_\beta(n,p,q_e)$  &  $ \max_{0\leq \theta \leq 1-p} H(\theta) - \theta H(p)-(1-\theta)H(\frac{p\theta}{1-\theta})$ & $\max_{0\leq \theta\leq 1}H((1-p)\theta)-\theta H(p)$ \\
    \hline
  \end{tabular}
\label{nonuniform_table1}
\end{table*}

\begin{figure*}[!t]
\centering\includegraphics[width=6in]{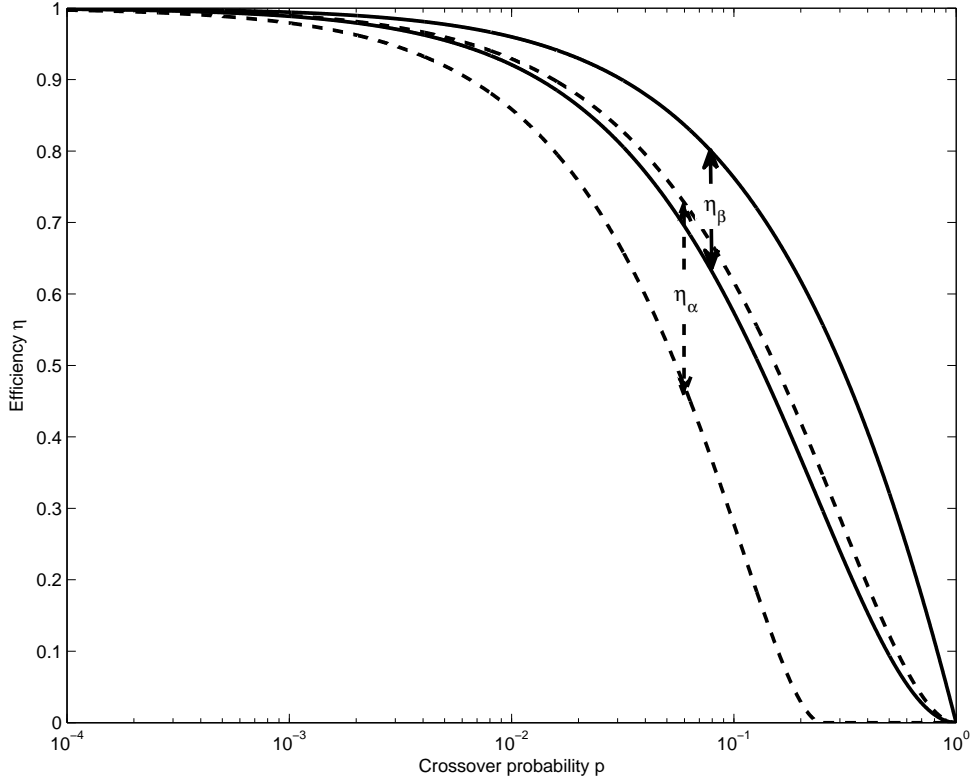}
\caption{Bounds of $\lim_{n\rightarrow\infty}\eta_\alpha(n,p,q_e)$
and $\lim_{n\rightarrow \infty}\eta_\beta(n,p,q_e)$.} \label{fig_asymp}
\end{figure*}

\begin{Theorem}[Upper bound]\label{theorem_asym4}
Given $0<p, q_e<1$, we have
\begin{eqnarray*}
\lim_{n\rightarrow \infty}\eta_\beta(n,p,q_e)&\leq& \max_{0\leq \theta\leq 1}H((1-p)\theta)-\theta H(p)\\
&=&H(\frac{1}{2^{s(p)}+1})+\frac{s(p)}{2^{s(p)}+1},\end{eqnarray*}
with $s(p)=H(p)/(1-p)$.
\end{Theorem}

\proof Here we use the same notations as above. Similar as the proof in Theorem~\ref{theorem_asym2}, given $(n,p,q_e)$, the maximal number of codewords  is
\begin{eqnarray*}
 B_\beta(n,p,q_e)&\leq &1+\sum_{w=\overline{h}(0)+1}^{n}\frac{{\nchoosek{n}{w-t_{\downarrow}(w)}}}{{\nchoosek{w}{ t_{\downarrow}(w)}}}\\
   &=& \sum_{w=\overline{h}(0)}^{n}\frac{{\nchoosek{n}{ w-t_{\downarrow}(w)}}}{{\nchoosek{w}{ t_{\downarrow}(w)}}}\\
   &\leq & \max_{w=0}^n  n \frac{{\nchoosek{n}{ w-t_{\downarrow}(w)}}}{{\nchoosek{w}{ t_{\downarrow}(w)}}}.
\end{eqnarray*}

As $n\rightarrow\infty$, we have
\begin{eqnarray*}
&&\eta_\beta(n,p,q_e)\\
&=& \frac{1}{n}\log_2 B_\beta(n,p,q_e)\\
&\leq & \frac{1}{n}\log_2 \max_{0\leq \theta\leq 1} n\frac{2^{H((1-\gamma)\theta+\delta)n}}{2^{(H(\gamma)\theta-\delta) n}}\\
&=& \max_{0\leq \theta\leq 1}H((1-\gamma)\theta)-\theta H(\gamma)+2\delta+\frac{1}{n}\log n\\
&=& \max_{0\leq \theta\leq 1}H((1-\gamma)\theta)-\theta H(\gamma).
\end{eqnarray*}

Note that when $\theta<\epsilon$ for small $\epsilon$, we have
$$H((1-\gamma)\theta)-\theta H(\gamma)\sim 0.$$
So we can ignore the edge effect. That implies that we can write
$$p-\delta\leq \gamma \leq p+\delta,$$
for any $\theta$ with $0\leq \theta\leq 1$.

Since for any fixed $\theta$ with $0\leq \theta \leq 1$, $H((1-\gamma)\theta)-\theta H(\gamma)$ is
a continuous function of $\gamma$. When $n\rightarrow \infty$, we have
$$\eta_\beta(n,p,q_e)\lesssim  \max_{0\leq \theta\leq 1}H((1-p)\theta)-\theta H(p),$$
which equals to
$$H(\frac{1}{2^{s(p)}+1})+\frac{s(p)}{2^{s(p)}+1},$$
with $s(p)=H(p)/(1-p)$.

This completes the proof.
\hfill\newQED

\subsection{Comparison of Asymptotic Performances}

Table \ref{nonuniform_table1} summarizes the analytic upper bounds and lower
bounds of $\lim_{n\rightarrow \infty}\eta_\alpha(n,p,q_e)$ and
$\lim_{n\rightarrow \infty}\eta_\beta(n,p,q_e)$ obtained in this section.
For the convenience of comparison, we plot them in
Figure~\ref{fig_asymp}. The dashed curves
represent the lower and upper bounds to
$\lim_{n\rightarrow\infty}\eta_\alpha(n,p,q_e)$, and the solid curves
represent the lower and upper bounds to
$\lim_{n\rightarrow\infty}\eta_\beta(n,p,q_e)$.
The gap between the bounds for the two
codes indicate the potential improvement in efficiency (code rate) by using
the nonuniform codes (compared to using uniform codes) when the
codeword length is large. We see that the upper bound in Theorem
\ref{theorem_asym4} is also the capacity of the Z-channel,
derived in \cite{Verdu1997}. It means that nonuniform codes may
be able to achieve the Z-channel capacity as $n$ becomes large,
while uniform codes cannot (here we assume that they have codewords of high weights and
worst-case performance is considered, so the constructions of uniform codes cannot achieve the capacity of Z-channel).

\section{Layered Codes Construction}
\label{section_construction1}

In \cite{Klove1995}, Kl{\o}ve summarized some constructions of uniform codes for correcting asymmetric errors.
The code of Kim and Freiman was the first one constructed for correcting multiple
asymmetric errors. Varshamov \cite{Varshamov1973} and Constrain and Rao \cite{Constantin1979}
presented some constructions based group theory. Later, Delsarte and Piret \cite{Delsarte1981} proposed
a construction based on `expurgating/puncturing' with some improvements given by Weber et al.\cite{Weber1988}.
It is natural for us to ask whether it is possible to construct nonuniform codes based on existing constructions of uniform codes.
In this section, we propose a general construction of nonuniform codes based on multiple layers. It shows that the sizes of the codes can be significantly increased by equalizing the reliability of all the codewords.

\subsection{Layered Codes}

Let us start from a simple example: Assume we want to construct a nonuniform code with codeword length $n=10$ and
$$t_{\downarrow}(w)=\left\{ \begin{array}{ll}
                               0 & \textrm{ for } w=0,\\
                              1 & \textrm{ for } 1\leq w\leq 5, \\
                              2 &  \textrm{ for } 6\leq w\leq 10.
                            \end{array}
\right.$$
In this case, how can we construct a nonuniform code efficiently? Intuitively,
we can divide all the codewords into two layers such that each layer corresponds to an individual uniform code, namely, we get a nonuniform code
\begin{eqnarray*}
  C &=& \{\mathbf{x}\in\{0,1\}^n| w(\mathbf{x})\leq 5, \mathbf{x}\in C_1\} \\
  && \bigcup \{\mathbf{x}\in\{0,1\}^n| w(\mathbf{x})\geq 6, \mathbf{x}\in C_2\},
\end{eqnarray*}
where $C_1$ is a uniform code correcting $1$ asymmetric error and $C_2$ is a uniform code correcting $2$ asymmetric errors.
So we can obtain a nonuniform code by combining multiple uniform codes, each of which
corrects a number of asymmetric errors. We call nonuniform codes constructed in this way as \emph{layered codes}.
However, the simple construction above has a problem -- due
to the interference of neighbor layers, the codewords at the bottom of the higher layer may violate our requirement of reliability, namely,
they cannot correct sufficient asymmetric errors. To solve this problem, we can construct a layered code in the following way: Let us first
construct a uniform code correcting $2$ asymmetric errors. Then we add more codewords into the code such that
\begin{enumerate}
\item The weights of these additional codewords are less than $4=6-t_{\downarrow}(6)$. This condition can guarantee that in the resulting nonuniform code all the codewords with weights at least $6$ can tolerate $2$ errors.
  \item These additional codewords are selected such that the codewords with weights at most $5$ can tolerate $1$ error.
\end{enumerate}

\subsection{Construction}

Generally, given a nondecreasing function $t_{\downarrow}$, we can get a nonuniform code with $t_{\downarrow}(n)$ layers by iterating the process
above. Based on this idea, given $n,t_{\downarrow}$, we construct layered codes as follows.

Let $k=t_{\downarrow}(n)$ and let $C_1,...,C_k$ be $k$ binary codes of codeword length $n$, where
$$C_1 \supset ...\supset C_k,$$
and for $1\leq t\leq
k$, the code $C_t$ can correct $t$ asymmetric errors.  Given $t_{\downarrow}$, we can construct a layered code $C$ such that
$$C=\{\mathbf{x}\in \{0,1\}^n|\mathbf{x}\in C_{t_{l}(w(\mathbf{x}))}\},$$ where
\begin{eqnarray*}
  t_l(w(\mathbf{x})) &=& t_{\downarrow}(\max R_{w(\mathbf{x})}) \\
   &=& t_{\downarrow}(\max
\{s| s-t_{\downarrow}(s)\leq w(\mathbf{x})\}).
\end{eqnarray*}

\begin{figure}[!t]
\centering
\includegraphics[width=3.6in]{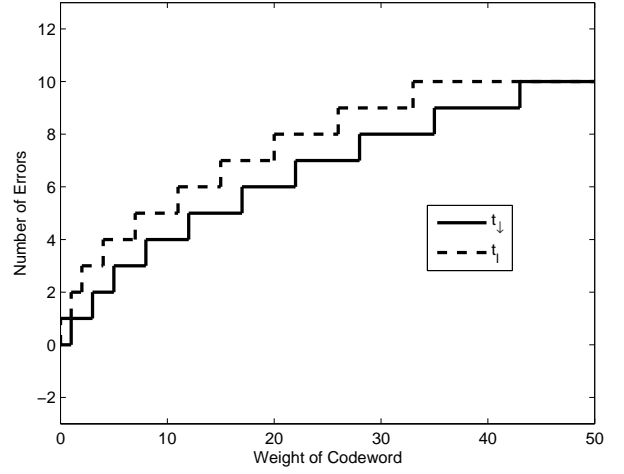}
\caption{A demonstration of function $t_{\downarrow}$ and $t_{l}$. }
\label{fig_layer}
\end{figure}

We see that there is a shift of the layers (corresponding to the function $t_l$ and the function $t_{\downarrow}$), see Figure \ref{fig_layer}
as a demonstration. The following theorem shows that the construction above satisfies our requirements of nonuniform codes, i.e., it corrects $
t_{\downarrow}$ asymmetric errors.

\begin{Theorem} Let $C$ be a layered code based on the above construction, then for all $\mathbf{x}\in
C$, $\mathbf{x}$ can tolerate $t_{\downarrow}(w(\mathbf{x}))$ asymmetric
errors.
\end{Theorem}

\proof We prove that for all $\mathbf{x},\mathbf{y}\in C$ with
$\mathbf{x}\neq \mathbf{y}$,
$\mathcal{B}(\mathbf{x})\bigcap
\mathcal{B}(\mathbf{y})=\phi$. W.l.o.g., we assume
$w(\mathbf{x})\geq w(\mathbf{y})$.

If $w(\mathbf{x})-t_{\downarrow}(w(\mathbf{x}))>w(\mathbf{y})$, the conclusion
is true.

If $w(\mathbf{x})-t_{\downarrow}(w(\mathbf{x}))\leq w(\mathbf{y})$ and
$w(\mathbf{x})\geq w(\mathbf{y})$, then $\mathbf{x},\mathbf{y} \in C_{t_{l}(w(\mathbf{y}))}$.
That means there does not exist a word $\mathbf{z} \in \{0,1\}^n$ such that
$\mathbf{x},\mathbf{y}\geq \mathbf{z}$ and $N(\mathbf{x},\mathbf{z})\leq t_l(w(\mathbf{y}))$ and
$N(\mathbf{y},\mathbf{z})\leq t_l(w(\mathbf{y}))$.  Since $w(\mathbf{x})-t_{\downarrow}(w(\mathbf{x}))\leq w(\mathbf{y})$,
according to the definition of $t_l$, it is easy to get $t_l(w(\mathbf{y}))\geq t_{\downarrow}(w(\mathbf{x}))\geq t_{\downarrow}(w(\mathbf{y}))$.
So there does not exist a word $\mathbf{z} \in \{0,1\}^n$ such that
$\mathbf{x},\mathbf{y}\geq \mathbf{z}$ and $N(\mathbf{x},\mathbf{z})\leq t_{\downarrow}(w(\mathbf{x}))$ and
$N(\mathbf{y},\mathbf{z})\leq t_{\downarrow}(w(\mathbf{y}))$, namely, $\mathcal{B}(\mathbf{x})\bigcap
\mathcal{B}(\mathbf{y})=\phi$.

This completes the proof. \hfill\newQED

We see that the constructions of layered codes are based on
the provided group of codes $C_1,...,C_{k}$ such that $ C_1 \supset C_2 \supset...\supset C_k$ and for $1\leq t\leq
k$, and the code $C_t$ corrects $t$ asymmetric errors. Examples of such codes include Varshamov codes \cite{Varshamov1973},
BCH codes, etc.

The construction of Varshamov codes can be described as follows: Let $\alpha_1,\alpha_2,...,\alpha_n$ be distinct nonzero
elements of $F_q$, and let
$\alpha:=(\alpha_1,\alpha_2,...,\alpha_n)$. For
$\mathbf{x}=(x_1,x_2,...,x_n)\in \{0,1\}^n$, let
$\mathbf{x}\alpha=(x_1\alpha_1,x_2\alpha_2,...,x_n\alpha_n)$. For
$g_1,g_2,...,g_t\in F_q$ and $1\leq t\leq k$, let
$$C_t:=\{\mathbf{x}\in\{0,1\}^n| \sigma_l(\mathbf{x}\alpha)=g_l \textrm{ for } 1\leq l\leq t\},$$
where the elementary symmetric
function $\sigma_l(\mathbf{u})$  for $l\geq 0$ are defined by
$$ \prod_{i=1}^r (z+u_i)=\sum_{l=0}^{\infty} \sigma_l(\mathbf{u})z^{r-l}.$$
Then $C_t$ can correct $t$ asymmetric errors (for $1\le t \le k$),
and $ C_1 \supset C_2 \supset ... \supset C_k$.

Such a group of codes can also be constructed by BCH codes: Let
$(\alpha_0, \alpha_1, ...,
\alpha_{n-1})$ be $n$ distinct nonzero elements of $G_{2^m}$ with $n=2^m-1$. For
$1\leq t\leq k$, let
$$C_t:=\{\mathbf{x}\in\{0,1\}^n| \sum_{i=1}^n x_i \alpha_i^{(2l-1)}=0 \textrm{ for } 1\leq l\leq t\}.$$

\subsection{Decoding Algorithm}

Assume $\mathbf{x}$ is a codeword in $C_t$ and
$\mathbf{y}=\mathbf{x}+\mathbf{e}$ is a received erroneous word with error vector $e$, then there is an efficient algorithm to decode
$\mathbf{y}$ into a codeword, which is denoted by
$D_t(\mathbf{y})$. If $\mathbf{y}$ has at most $t$ asymmetric
errors, then $D_t(\mathbf{y})=\mathbf{x}$. We show that the layered codes proposed above also have an
efficient decoding algorithm if $D_t(\cdot)$ (for $1 \le t \le k$)
are provided and efficient.

\begin{Theorem}
Let $C$ be a layered code based on the above construction, and let $\mathbf{y}=\mathbf{x}+\mathbf{e}$ be a received word
such that $\mathbf{x}\in C$ and $|e|\leq t_{\downarrow}(w(\mathbf{x}))$. To recover $\mathbf{x}$ from $\mathbf{y}$, we
enumerate the integers in $[t_l(w(\mathbf{y})),t_l(w(\mathbf{y})+t_l(w(\mathbf{y})))]$.
If we can find an integer $t$ such that $D_t(\mathbf{y})\in C$ and $N(D_t(\mathbf{y}),\mathbf{y})\leq t_{\downarrow}(w(D_t(\mathbf{y})))$,
then $D_t(\mathbf{y})=\mathbf{x}$.\end{Theorem}

\proof If we let $t=t_{\downarrow}(w(\mathbf{x}))$, then we can get that $t$
satisfies the conditions and $D_t(\mathbf{y})=\mathbf{x}$. So
such $t$ exists.

Now we only need to prove that once there exists $t$ satisfying
the conditions in the theorem, we have
$D_t(\mathbf{y})=\mathbf{x}$. We prove this by contradiction.
Assume there exists $t$ satisfying the conditions but
$\mathbf{z}=D_t(\mathbf{y})\neq \mathbf{x}$. Then
$N(\mathbf{z},\mathbf{y})\leq t_{\downarrow}(w(\mathbf{z}))$.
Since we also have $N(\mathbf{x},\mathbf{y})\leq t(w(\mathbf{x}))$,
$\mathcal{B}(\mathbf{x})\bigcap \mathcal{B}(\mathbf{z})\neq \phi$, which contradicts
the property of the layered codes.

This completes the proof.\hfill\newQED

In the above method, to decode an erroneous word
$\mathbf{y}$, we can check all the integers between
$t_l(w(\mathbf{y}))$ and $t_l(w(\mathbf{y})+t_l(w(\mathbf{y})))$ to
find the value of $t$. Once we find the integer $t$ satisfying the
conditions in the theorem, we can decode $\mathbf{y}$ into
$D_t(\mathbf{y})$ directly. (Note that the length of the interval for $t$, namely
$t_l(w(\mathbf{y})+t_l(w(\mathbf{y})))-t_l(w(\mathbf{y}))$, is
normally much smaller than $w(\mathbf{y})$. It is approximately
$\frac{p^2}{(1-p)^2}w(\mathbf{y})$ for i.i.d. errors when $w(\mathbf{y})$ is large.) We see that this decoding
process is efficient if $D_t(.)$ is efficient for $1\leq t\leq k$.

\subsection{Layered vs.Uniform}

Typically, nonlinear codes, like Varshamov codes are superior to BCH codes. But it is still not well-known how to estimate the sizes of Varshamov codes and their weight distributions.
To compare uniform constructions and nonuniform constructions for correcting asymmetric errors,
we focus on BCH codes, namely, we compare normal BCH codes with layered BCH codes. Here, we consider
i.i.d. errors, and we assume that the codeword length is $n=255$, the crossover probability is $p$ and
the maximal tolerated error probability is $q_e$.

\begin{table}
    \centering
    \caption{BCH codes with codeword length $255$  \cite{Fujiwara06}}
    \renewcommand{\arraystretch}{1.25}
  \begin{tabular}{|lllllll|}
    \hline
     n& k & t & \quad\quad\quad & n & k & t\\
    \hline
    255 & 247 & 1 && 255 & 115 & 21\\
    255 & 239 & 2 && 255 & 107 & 22 \\
    255 & 231 & 3 && 255 & 99 & 23\\
    255 & 223 & 4 && 255 & 91 & 25\\
    255 & 215 & 5 && 255 & 87 & 26\\
    255 & 207 & 6 && 255 & 79 & 27 \\
    255 & 199 & 7 && 255 & 71 & 29\\
    255 & 191 & 8 && 255 & 63 & 30\\
    255 & 187 & 9 && 255 & 55 & 31\\
    255 & 179 & 10 && 255 & 47 & 42\\
    255 & 171 & 11 && 255 & 45 & 43\\
    255 & 163 & 12 && 255 & 37 & 45\\
    255 & 155 & 13 && 255 & 29 & 47\\
    255 & 147 & 14 && 255 & 21 & 55\\
    255 & 139 & 15 && 255 & 13 & 59\\
    255 & 131 & 18 && 255 & 9 & 63\\
    255 & 123 & 19 && & & \\
    \hline
  \end{tabular}
  \label{nonuniform_table2}
\end{table}

Table \ref{nonuniform_table2} shows the relations between the dimension $k$ and the number of errors $t$ that can be corrected in BCH codes when $n=255$.
According to \cite{Macwilliams77}, many BCH codes have approximated binomial weight distribution.
So given an $(255, k, t)$ BCH code, the number of codewords of weight $i$ is approximately
$$b_i\sim 2^k\frac{{\nchoosek{n}{ i}}}{2^n}.$$

For a normal BCH code, it has to correct $t$ errors with
$$t=\min\{s\in N| \sum_{i=0}^s {\nchoosek{n}{ i}}p^i(1-p)^{n-i}\geq 1-q_e\},$$
then it has $2^k$ codewords where $k$  can be obtained from table \ref{nonuniform_table2} based on the value of $t$.

For a layered BCH code,
the codewords with Hamming weight $w$ have
to correct $t_{\downarrow}(w)$ asymmetric errors such that
$$t_{\downarrow}(w)=\min\{s\in N| \sum_{i=0}^s {\nchoosek{w}{ i}} p^i(1-p)^{w-i}\geq 1-q_e\},$$
for all $0\leq w\leq n$. Based on the approximated weight distribution of BCH codes, the number of codewords in a layered BCH codes
can be estimated by summing up the numbers of codewords with different weights.

\begin{figure}[!t]
\centering
\includegraphics[width=3.6in]{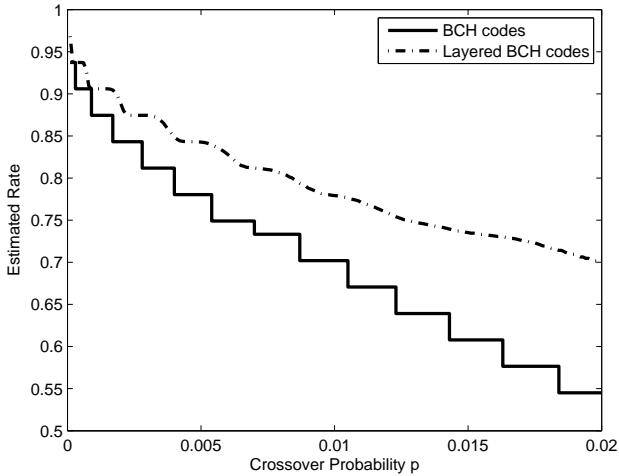}
\caption{The estimated rates of BCH codes and layered BCH codes when $n=255, q_e=10^{-4}$. }
\label{fig_BCHrate}
\end{figure}

Figure \ref{fig_BCHrate} plots
the estimated rates of BCH codes and layered BCH codes for different $p$ when $n=255$ and $q_e=10^{-4}$.
Here, for a code $C$, let $\#C$ be the number of codewords, then the rate of $C$ is defined as $\frac{\log_2(\#C)}{n}$. From this figure,
we see that under the same parameters $(n,p,q_e)$,
the rates of layered BCH codes are much higher than those of BCH codes. By constructing nonuniform codes instead of uniform codes, the code rate can be significantly increased.
Comparing Figure \ref{fig_BCHrate} with Figure \ref{fig_upbound2}, it can be seen that the rates
of layered BCH codes are very close to the upper bounds of uniform codes. It implies that we can gain more by considering nonuniform codes rather than nonlinear
uniform codes.

\section{Flipping Codes Construction}
\label{section_construction2}

Many nonlinear codes
designed to correct asymmetric
errors like Varshamov codes are superior to linear codes. However, they do not yet have efficient encoding algorithms, namely, it
is not easy to find an efficient encoding function $f:
\{0,1\}^k\rightarrow C$ with $k\backsimeq \lfloor\log |C|\rfloor
$.  In this section, we focus on the approach of designing nonuniform codes for
asymmetric errors with efficient encoding schemes, by utilizing the well-studied linear codes.

A simple method is that we can use a linear code to correct $t_{\downarrow}(n)$ asymmetric errors directly, but this method is inefficient not only
because the decoding sphere for symmetric errors is greater than
the sphere for asymmetric errors (and therefore an overkill), but
also because for low-weight codewords, the number of asymmetric
errors they need to correct can be much smaller than $t_{\downarrow}(n)$.

Our idea is to build a \emph{flipping code} that uses only low-weight
codewords (specifically, codewords of Hamming weight no more than
$\sim\frac{n}{2}$), because they need to correct fewer
asymmetric errors and therefore can increase the code's rate.
In the rest of this section, we present two different constructions.

\subsection{First Construction}

First, we construct a linear code $C$ (like BCH codes) of length $n$ with generator
matrix $G$ that corrects $t_{\downarrow}(\lfloor\frac{n}{2}\rfloor)$ symmetric
errors. Assume the dimension of the code is $k$. For any binary
message $\mathbf{u}\in \{0,1\}^k$, we can map it to a codeword
$\mathbf{x}$ in $C$ such that $\mathbf{x}=\mathbf{u}G$.  Next, let $\overline{\mathbf{x}}$
denote a word obtained by flipping all the bits in $\mathbf{x}$ such that if
$x_i=0$ then $\overline{x}_i=1$ and if $x_i=1$ then $\overline{x}_i=0$; and let
$\mathbf{y}$ denote the final codeword corresponding to $\mathbf{u}$. We check whether $w(\mathbf{x})\leq \lfloor\frac{n}{2}\rfloor$ and construct $\mathbf{y}$ in the following way:
$$\mathbf{y}=\left\{\begin{array}{cc}
           \mathbf{x}00...0 & \textrm{ if } w(\mathbf{x})\leq \lfloor\frac{n}{2}\rfloor, \\
           \mathbf{\overline{x}}11...1 & \textrm{ otherwise.}
         \end{array}\right.
$$
Here, the auxiliary bits ($0$s or $1$s) are added to distinguish that whether $\mathbf{x}$ has been flipped or not, and they
form a repetition code to tolerate errors.

The corresponding decoding process is straightforward: Assume we received a word $\mathbf{y'}$.
If there is at least one $1$
in the auxiliary bits, then we ``flip'' the word by
changing all $0$s to $1$s and all $1$s to $0$s; otherwise, we keep the
word unchanged. Then we apply the decoding scheme of the code $C$ to
the first $n$ bits of the word. Finally, the message $\mathbf{u}$ can be successfully decoded if
$\mathbf{y'}$ has at most $t_{\downarrow}(\lfloor\frac{n}{2}\rfloor)$ errors in the first $n$ bits.

\subsection{Second Construction}
In the previous construction, several
auxiliary bits are needed to protect one bit
of information, which
is not very efficient. Here we try to move this bit into the information part of the codewords in $C$.
This motivates us to give the following construction.

Let $C$ be a systematic linear code with length $n$ that corrects $t'$ symmetric errors (we
will specify $t'$ later).
Assume the dimension of the code is $k$. Now, for any binary message $\mathbf{u}\in \{0,1\}^{k-1}$ of length $k-1$,
we get $\mathbf{u'}=0\mathbf{u}$ by adding one bit $0$ in front of $\mathbf{u}$.  Then we can map $\mathbf{u'}$ to a codeword $\mathbf{x}$ in
$C$ such that $$\mathbf{x}=(0\mathbf{u})G=0\mathbf{uv},$$ where $G$ is the generator matrix of $C$ in systematic form and the length of $\mathbf{v}$ is $n-k$. Let $\mathbf{\alpha}$ be a codeword in $C$ such that the first bit $\alpha_1=1$
and its weight is the maximal one among all the codeword in $C$, i.e.,
$$\mathbf{\alpha}=\arg \max_{\mathbf{x}\in C, x_1=1} w(\mathbf{x}).$$
Generally, $w(\mathbf{\alpha})$ is very close to $n$. For example, in any primite BCH code of length $255$,
$\mathbf{\alpha}$ is the all-one vector; also we can construct LDPC codes that include the all-one vector as long as their parity-check matrices have even number of ones in each column.
In order to reduce the weights of the codewords, we use the
following operations: Calculate the relative weight $$w(\mathbf{x}|\mathbf{\alpha})=|\{1\leq i\leq n|x_i=1, \alpha_i=1\}|.$$
Then we get the final codeword
$$\mathbf{y}=\left\{\begin{array}{cc}
           \mathbf{x}+\mathbf{\alpha} & \textrm{ if } w(\mathbf{x}|\mathbf{\alpha})>\frac{w(\mathbf{\alpha})}{2},\\
           \mathbf{x} & \textrm{ otherwise,}
         \end{array}\right.$$
where $+$ is the binary sum, so $\mathbf{x}+\mathbf{\alpha}$ is to flip the bits in $\mathbf{x}$ corresponding the ones in $\mathbf{\alpha}$.
So far, we see that the maximal weight for $\mathbf{y}$ is $\lfloor n-\frac{w(\mathbf{\alpha})}{2}\rfloor$. That means we need to
select $t'$ such that $$t'=t_{\downarrow}(\lfloor n-\frac{w(\mathbf{\alpha})}{2}\rfloor).$$
For many linear codes, $\alpha$ is the all-one vector, so $t'=t_{\downarrow}(\lfloor \frac{n}{2}\rfloor).$

In the above encoding process, for different binary messages, they have different codewords. And for any codeword $\mathbf{y}$, we have $\mathbf{y}\in C$.
That is because either $\mathbf{y}=\mathbf{x}$ or $\mathbf{y}=\mathbf{x}+\mathbf{\alpha}$, where both $\mathbf{x}$ and $\mathbf{\alpha}$ are codewords in $C$ and $C$ is a linear code.
So the resulting flipping code is a subset of code $C$.

The decoding process is very simple: Given the received word $\mathbf{y'}=\mathbf{y}+\mathbf{e}$, we can always get $\mathbf{y}$ by applying the decoding scheme of the linear code $C$ if $|\mathbf{e}|\leq t'$.
If $y_1=1$, that means $\mathbf{x}$ has been flipped based on $\mathbf{\alpha}$, so we have $\mathbf{x}=\mathbf{y}+\mathbf{\alpha}$; otherwise, $\mathbf{x}=\mathbf{y}$.
Then the initial message $\mathbf{u}=x_2x_3...x_k$.

We see that the second construction is a little more efficient than the first one, by moving the bit that indicates flips from
the outside of a codeword (of an error-correcting code) to the inside. Here is an example of the second construction: Let $C$
be the $(7,4)$ Hamming code, which is able to correct single-bit errors. The generating matrix
of the $(7,4)$ Hamming code is
$$G=\left(
      \begin{array}{ccccccc}
        1 & 0 & 0 & 0 & 1 & 1 & 0 \\
        0 & 1 & 0 & 0 & 1 & 0 & 1 \\
        0 & 0 & 1 & 0 & 0 & 1 & 1 \\
        0 & 0 & 0 & 1 & 1 & 1 & 1 \\
      \end{array}
    \right).
$$
Here we have $t'=1$ and $k=4$. Assume the binary message is $\mathbf{u}=011$, then
we have $\mathbf{x}=(0\mathbf{u})G=0011100$. It is easy to see that $\mathbf{\alpha}$ is the all-one codeword, i.e.,
$\mathbf{\alpha}=1111111$. In this case, $w(\mathbf{x}|\mathbf{\alpha})<=\frac{w(\mathbf{\alpha})}{2}$, so the final codeword $\mathbf{y}=0011100$.
Assume the binary message is $\mathbf{u}=110$, then we have $\mathbf{x}=(0\mathbf{u})G=0110110$. In this case,
$w(\mathbf{x}|\mathbf{\alpha})>\frac{w(\mathbf{\alpha})}{2}$, so the final codeword $\mathbf{y}=\mathbf{x}+\mathbf{\alpha}=1001001$.

Assume the received word is $\mathbf{y'}=0001001$. By applying the decoding algorithm of Hamming codes,
we get $\mathbf{y}=1001001$. Since $y_1=1$, we have $\mathbf{x}=\mathbf{y}+\mathbf{\alpha}$, and as a result, $\mathbf{u}=110$.

\subsection{Flipping vs.Layered}

When $n$ is sufficiently large, the flipping codes above
become nearly as efficient (in terms of code rate) as a linear codes correcting
$t_{\downarrow}(\lfloor\frac{n}{2}\rfloor)$ symmetric errors.  It is much more
efficient than designing a linear code correcting $t_{\downarrow}(n)$ symmetric
errors. Note that when $n$ is large and $p$ is small, these codes can have very good performance on code rate.
That is because when $n$ is sufficiently large, the rate of an optimal nonuniform code is dominated
by the codewords with the same Hamming weight $w_d$ ($\leq \frac{n}{2}$), and
$w_d$ approaches $\frac{n}{2}$ as $p$ gets close to $0$. We can intuitively understand it based on two facts when $n$ is sufficiently large:
(1) There are at most $n 2^{n(H(\frac{w_d}{n})+\delta)}$ codewords in this optimal nonuniform code.
(2) When $p$ becomes small, we can get a nonuniform code with at least $2^{n(1-\delta)}$ codewords.
So when $n$ is sufficiently large and $p$ is small, we have $w_d\rightarrow \frac{n}{2}$. Hence, an optimal nonuniform code has almost the same asymptotic performance with an optimal
weight-bounded code (Hamming weight is at most n/2) that corrects $t_{\downarrow}(n/2)$ asymmetric errors.

\begin{figure}[!t]
\centering
\includegraphics[width=3.6in]{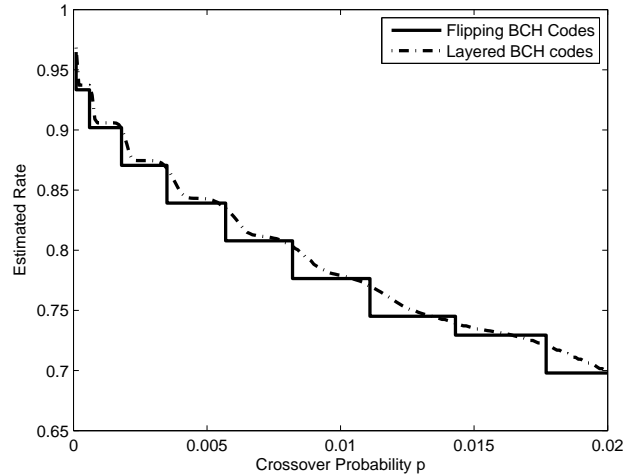}
\caption{The estimated rates of flipping/layered BCH codes when $n=255, q_e=10^{-4}$. }
\label{fig_flippingrate}
\end{figure}

Let us consider a flipping BCH code based on the second construction. Similar as the previous section, we assume
that the codeword length is $n=255$ and the number of codewords with weight $i$ can be approximated by
$$ 2^k\frac{{\nchoosek{n}{ i}}}{2^n},$$
where $k$ is the dimension of the code. Figure \ref{fig_flippingrate} compares the estimated rates of flipping BCH codes
and those of layered BCH codes when $n=255$ and $q_e=10^{-4}$. Surprisingly, the flipping BCH codes achieves almost the same rates
as layered BCH codes. Note that, for the layered codes, we are able to further improve the efficiency (rates) by replacing BCH codes with Varshamov codes, i.e., based on layered Varshamov codes.

\section{Extension to Binary Asymmetric Channels}
\label{section_general}

In the previous sections, we have introduced and studied nonuniform codes for Z-channels.
The concept of nonuniform codes can be extended from Z-channels to general binary asymmetric channels, where the error probability
from $0$ to $1$ is smaller than the error probability from $1$ to $0$ but it may not be ignorable. In this case, we are able to construct nonuniform codes
correcting a big number of $1\rightarrow 0$ errors and a small number of $0\rightarrow 1$ errors. Such codes can be used in flash memories or phase change memories, where
the change in data has an asymmetric property. For
example, the stored data in flash memories is represented by
the voltage levels of transistors, which drift in one direction because
of charge leakage. In phase change memories, another class of nonvolatile
memories, the stored data is determined by the electrical
resistance of the cells, which also drifts due to thermally
activated crystallization of the amorphous material. This asymmetric property will introduce more $1\rightarrow 0$ errors than $0\rightarrow 1$ errors after a long duration.

In this section, we first investigate binary asymmetric channels where the probability from $0$ to $1$ is much smaller than that
from $1$ to $0$, namely, $p_{\uparrow}\ll p_{\downarrow}$, but $p_{\uparrow}$ is not ignorable.  In this case,
we can let $t_{\uparrow}$ be a constant function.
Later, we consider general binary asymmetric channels, where $t_{\uparrow}$ can be an arbitrary nonincreasing step function.

\subsection{$t_{\uparrow}$ Is a Constant Function}

We show that if $t_{\uparrow}$ is a constant function, then correcting $[t_{\downarrow}, t_{\uparrow}]$ errors is
equivalent to correcting $t_{\downarrow}+t_{\uparrow}$ asymmetric errors, where $t_{\downarrow}$ can be an arbitrary step functions on $\{0,1,...,n\}$.

\begin{Theorem} \label{theorem_asymmetric1} Let $t_{\uparrow}$ be a constant function, a code $C$ is a nonuniform code correcting $[t_{\downarrow}, t_{\uparrow}]$ errors
if and only if it is a nonuniform code correcting $t_{\downarrow}+t_{\uparrow}$ asymmetric errors.
\end{Theorem}

\proof
1) We first show that if $C$ is a nonuniform code correcting $[t_{\downarrow}, t_{\uparrow}]$ errors where $t_{\uparrow}$ is a constant function,
then it can correct $t_{\downarrow}+t_{\uparrow}$ asymmetric errors.
We need to prove that there does not exists a pair of codewords $\mathbf{x},\mathbf{y}\in C$ such that
$$N(\mathbf{x},\mathbf{y})\leq t_{\downarrow}(w(\mathbf{x}))+t_{\uparrow},$$
$$N(\mathbf{y},\mathbf{x})\leq t_{\downarrow}(w(\mathbf{y}))+t_{\uparrow},$$
where
$$N(\mathbf{x},\mathbf{y}) \triangleq |\{i:x_i=1,y_i=0\}|.$$

\begin{figure}[!t]
\centering
\includegraphics[width=3.2in]{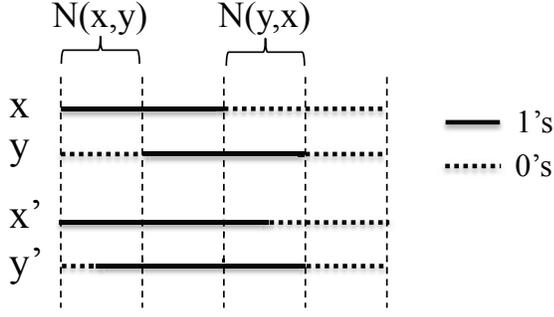}
\caption{A demonstration of $\mathbf{x},\mathbf{y}, \mathbf{x}',\mathbf{y}'$.}
\label{fig_general1}
\end{figure}

Let us prove it by contradiction. Assume that their exists a pair of codewords $\mathbf{x},\mathbf{y}$ that satisfy the inequalities above.
By adding at most $t_{\uparrow}$ $0\rightarrow 1$ errors, we get a vector $\mathbf{x'}$ from $\mathbf{x}$ such that
the Hamming distance between $\mathbf{x'}$ and $\mathbf{y}$ is minimized; also we get a vector $\mathbf{y'}$ from $\mathbf{y}$
such that the Hamming distance between $\mathbf{y'}$ and $\mathbf{x}$ is minimized. In this case, we only need to show that
$$N(\mathbf{x'},\mathbf{y'})\leq t_{\downarrow}(w(\mathbf{x})), N(\mathbf{y'},\mathbf{x'})\leq t_{\downarrow}(w(\mathbf{y})),$$
which contradicts with our assumption that $C$ can correct $[t_{\downarrow}, t_{\uparrow}]$ errors. The intuitive way of understanding $\mathbf{x}', \mathbf{y}'$
is shown in Figure \ref{fig_general1}. In the figure, we present each vector as a line, in which the solid part is for $1$s and the dashed part is for $0$s.

If $N(\mathbf{x'},\mathbf{x})<t_{\uparrow}$ and $N(\mathbf{y'},\mathbf{y})<t_{\uparrow}$,
then $${x}'_i=\max({x}_i, {y}_i)={y}'_i,$$ so
$\mathbf{x'}=\mathbf{y'}$. The statement is true.

If $N(\mathbf{x'},\mathbf{x})<t_{\uparrow}$ and $N(\mathbf{y'},\mathbf{y})=t_{\uparrow}$, then
$\mathbf{y'}\leq \mathbf{x'}$. In this case,
$$N(\mathbf{x'},\mathbf{y'})\leq N(\mathbf{x},\mathbf{y})-t_{\uparrow}\leq t_{\downarrow}(w(\mathbf{x})).$$
We get the statement.

Similarly, if $N(\mathbf{y'},\mathbf{y})<t_{\uparrow}$ and $N(\mathbf{x'}, \mathbf{x})=t_{\uparrow}$, we have
$\mathbf{x'}\leq \mathbf{y'}$ and
$$N(\mathbf{y'},\mathbf{x'})\leq N(\mathbf{y},\mathbf{x})-t_{\uparrow}\leq t_{\downarrow}(w(\mathbf{y})).$$

If $N(\mathbf{x'},\mathbf{x})=t_{\uparrow}$ and $N(\mathbf{y'},\mathbf{y})=t_{\uparrow}$, we can get
$$N(\mathbf{x'},\mathbf{y'})\leq N(\mathbf{x},\mathbf{y})-t_{\uparrow}\leq t_{\downarrow}(w(\mathbf{x})),$$
$$N(\mathbf{y'},\mathbf{x'})\leq N(\mathbf{y},\mathbf{x})-t_{\uparrow}\leq t_{\downarrow}(w(\mathbf{y})).$$

Based on the discussions above, we can conclude that if $C$ is a nonuniform code correcting $[t_{\downarrow}, t_{\uparrow}]$ errors where $t_{\uparrow}$ is a constant function,
then it is also a nonuniform code correcting $t_{\downarrow}+t_{\uparrow}$ asymmetric errors.

2) We show that if $C$ is a nonuniform codes correcting $t_{\downarrow}+t_{\uparrow}$ asymmetric errors where $t_{\uparrow}$ is a constant function, then
it is also a nonuniform code correcting $[t_{\downarrow},t_{\uparrow}]$ errors.
That means for any $\mathbf{x},\mathbf{y}\in C$, there does not exist a vector $\mathbf{v}$ such that
$$N(\mathbf{v},\mathbf{x})\leq t_{\uparrow}, \quad N(\mathbf{x},\mathbf{v})\leq t_{\downarrow}(w(\mathbf{x})),$$
$$N(\mathbf{v},\mathbf{y})\leq t_{\uparrow}, \quad N(\mathbf{y},\mathbf{v})\leq t_{\downarrow}(w(\mathbf{y})).$$

\begin{figure}[!t]
\centering
\includegraphics[width=3.2in]{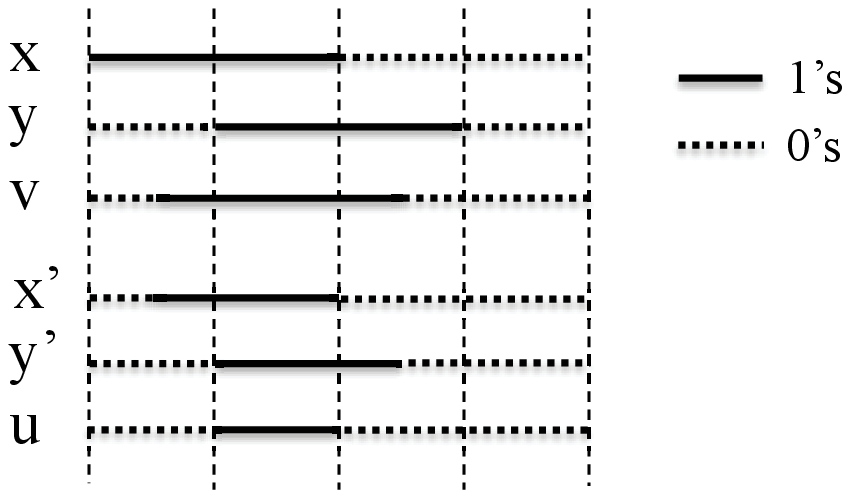}
\caption{A demonstration of $\mathbf{x},\mathbf{y}, \mathbf{x}',\mathbf{y}', \mathbf{v}, \mathbf{u}$.}
\label{fig_general2}
\end{figure}

Let us prove this by contradiction. We assume there exists a vector $\mathbf{v}$ satisfies the above conditions.
Now, we define a few vectors $\mathbf{x'}, \mathbf{y'}, \mathbf{u}$ such that
$${x'}_i=\min({x}_i, {v}_i) \quad \forall1\leq i\leq n,$$
$${y'}_i=\min({y}_i,{v}_i )\quad \forall1\leq i\leq n,$$
$${u}_i=\min({x}_i,{y}_i, {v}_i )\quad \forall1\leq i\leq n.$$
The intuitive way of understanding these vectors
is shown in Figure \ref{fig_general2}. In the figure, we present each vector as a line, in which the solid part is for $1$s and the dashed part is for $0$s.

Then
$$\mathbf{x'}\leq \mathbf{x},\mathbf{x'}\leq \mathbf{v}, N(\mathbf{x},\mathbf{x'})\leq t_{\downarrow}(w(\mathbf{x})), N(\mathbf{v},\mathbf{x'})\leq t_{\uparrow},$$
$$\mathbf{y'}\leq \mathbf{y},\mathbf{y'}\leq\mathbf{v}, N(\mathbf{y},\mathbf{y'})\leq t_{\downarrow}(w(\mathbf{y})), N(\mathbf{v},\mathbf{y'})\leq t_{\uparrow}.$$

Now we want to show that $$N(\mathbf{x},\mathbf{u})\leq t_{\downarrow}(w(\mathbf{x}))+ t_{\uparrow}.$$

Since $$N(\mathbf{x},\mathbf{u})\leq N(\mathbf{x},\mathbf{x'})+N(\mathbf{x'},\mathbf{u}),$$
we only to show that
$$N(\mathbf{x'},\mathbf{u})\leq t_{\uparrow}.$$

According to the definition of $\mathbf{u}$, it is easy to get that
\begin{eqnarray*}
 N(\mathbf{v},\mathbf{x'})+N(\mathbf{x'},\mathbf{u})
   &=& N(\mathbf{v},\mathbf{y'})+N(\mathbf{y'},\mathbf{u}) \\
   &\leq &  N(\mathbf{v},\mathbf{x'}) + N(\mathbf{v},\mathbf{y'})
\end{eqnarray*}
So $N(\mathbf{x'},\mathbf{u})\leq t_{\uparrow}$, which leads us to
$$N(\mathbf{x},\mathbf{u})\leq t_{\downarrow}(w(\mathbf{x}))+ t_{\uparrow}.$$

Similarly, we can also get
$$N(\mathbf{y},\mathbf{u})\leq t_{\downarrow}(w(\mathbf{y}))+ t_{\uparrow}.$$
In this case, $C$ is not a nonuniform codes correcting $t_{\downarrow}+t_{\uparrow}$ asymmetric errors, which
contradicts with our assumption.

Based on the discussions above, we can get the conclusion in the theorem.
\hfill \newQED

According to the above theorem, all our results for Z-channels, like upper bounds and constructions of nonuniform codes, can apply
to nonuniform codes correcting $[t_{\downarrow}, t_{\uparrow}]$ errors if $t_{\uparrow}$ is a constant function.

\subsection{$t_{\uparrow}$ Is a Nonincreasing Function}

Another case of binary asymmetric channel is that $p_{\uparrow}<p_{\downarrow}$ but $p_{\uparrow}$ is not much smaller than $p_{\downarrow}$.
In this case, it is not efficient to write $t_{\uparrow}$ as a constant function. Instead, we consider it as
a nonincreasing step function.

\begin{Theorem}
Let $t_{\downarrow}$ be a nondecreasing function and $t_{\uparrow}$ be a nonincreasing function.
A code $C$ is a nonuniform code correcting $[t_{\downarrow}, t_{\uparrow}]$ errors if it is
a nonuniform code correcting $t_{\downarrow}+\overline{t_{\uparrow}}$ asymmetric errors. Here, for all $0\leq w\leq n$,
$$\overline{t_{\uparrow}}(w)=t_{\uparrow}(\max\{s| t_{\uparrow}(s)+ s \leq w-t_{\downarrow}(w)\}).$$
\end{Theorem}

\proof Let $C$ be a nonuniform code correcting  $t_{\downarrow}+\overline{t_{\uparrow}}$ errors.
For any $\mathbf{x},\mathbf{y}\in C$, w.l.o.g, we assume
$w(\mathbf{x})\leq w(\mathbf{y})$. If $w(\mathbf{x})+t_{\uparrow}(w(\mathbf{x}))< w(\mathbf{y})-
t_{\downarrow}(w(\mathbf{y}))$, then there does not exist a vector $\mathbf{v}$ such that
$$N(\mathbf{v},\mathbf{x})\leq t_{\uparrow}, \quad N(\mathbf{x},\mathbf{v})\leq t_{\downarrow}(w(\mathbf{x})),$$
$$N(\mathbf{v},\mathbf{y})\leq t_{\uparrow}, \quad N(\mathbf{y},\mathbf{v})\leq t_{\downarrow}(w(\mathbf{y})).$$

If  $w(\mathbf{x})+t_{\uparrow}(w(\mathbf{x}))\geq  w(\mathbf{y})-
t_{\downarrow}(w(\mathbf{y}))$, according to the proof in Theorem~\ref{theorem_asymmetric1},
we can get that there does not exist a vector $\mathbf{v}$ such that
$$N(\mathbf{v},\mathbf{x})\leq t_{\uparrow}(w(\mathbf{x})),$$
$$ N(\mathbf{x},\mathbf{v})\leq t_{\downarrow}(w(\mathbf{x}))+\overline{t_{\uparrow}}(w(\mathbf{x}))-t_{\uparrow}(w(\mathbf{x}));$$
$$N(\mathbf{v},\mathbf{y})\leq t_{\uparrow}(w(\mathbf{x})),$$
$$ N(\mathbf{y},\mathbf{v})\leq t_{\downarrow}(w(\mathbf{y}))+\overline{t_{\uparrow}}(w(\mathbf{y}))-t_{\uparrow}(w(\mathbf{x})).$$

Since
$$\overline{t_{\uparrow}}(w(\mathbf{x}))-t_{\uparrow}(w(\mathbf{x}))\geq 0,$$
$$t_{\uparrow}(w(\mathbf{x}))\geq t_{\uparrow}(w(\mathbf{y})),$$
$$\overline{t_{\uparrow}}(w(\mathbf{y}))\geq t_{\uparrow}(w(\mathbf{x})),$$
we can get that there does not exist a vector $\mathbf{v}$ such that
$$N(\mathbf{v},\mathbf{x})\leq t_{\uparrow}, \quad N(\mathbf{x},\mathbf{v})\leq t_{\downarrow}(w(\mathbf{x})),$$
$$N(\mathbf{v},\mathbf{y})\leq t_{\uparrow}, \quad N(\mathbf{y},\mathbf{v})\leq t_{\downarrow}(w(\mathbf{y})).$$

Finally, we conclude that $C$ is a nonuniform code correcting $[t_{\downarrow},t_{\uparrow}]$ errors.
\hfill\newQED

According to the above theorem, we can convert the problem of constructing a nonuniform codes for an arbitrary binary asymmetric channel
to the problem of constructing a nonuniform correcting only $1\rightarrow 0$ errors. Note that this conversion results in
a little loss of code efficiency, but typically it is very small. Both layered codes and flipping codes can be applied for correcting errors in binary asymmetric channels. A little point to notice is that $t_{\downarrow}+\overline{t_{\uparrow}}$ might not be a strict nondecreasing function of codeword weight. In this case,
we can find a nondecreasing function $t_{h}$ which is slightly larger than $t_{\downarrow}+\overline{t_{\uparrow}}$, and  construct a nonuniform code correcting $t_h$ asymmetric errors.

When we apply flipping codes for correcting errors in binary asymmetric channels, we do not have to specify $t_{\downarrow}$ and $t_{\uparrow}$ separately.
For example, assume that i.i.d. errors are considered. If the maximal tolerated error probability is $q_e$, then given
a codeword of weight $w$, it has to tolerate total $t_{f}(w)$ errors. For $0\leq w\leq n$, $t_f(w)$ can be obtained by calculating the minimal integer $t$ such that
$$\sum_{i=0}^{t}\sum_{j=0}^{t-i}{\nchoosek{w}{ i}}{\nchoosek{n-w}{ j}} p_{\downarrow}^i(1-p_{\downarrow})^{w-i}p_{\uparrow}^j(1-p_{\uparrow})^{(n-w-j)}$$
$$\geq 1-q_e.$$
To construct a flipping code, we only need to find a linear code such that it corrects
$t_{f}(\lfloor n-\frac{\alpha}{2}\rfloor)$
symmetric errors, where $\alpha$ is the codeword with the maximum weight in the linear code.

\begin{Theorem} Let $t_{\downarrow}$ be a nondecreasing function and $t_{\uparrow}$ be a nonincreasing function. If a code $C$ is a nonuniform code correcting $[t_{\downarrow},t_{\uparrow}]$ errors,
then it corrects $t_{\downarrow}+\underline{t_{\uparrow}}$ asymmetric errors. Here,
$$\underline{t_{\uparrow}}(w)=t_{\uparrow}(\min\{s| s-t_{\uparrow}(s)- t_{\downarrow}(s)\leq w\}).$$
\end{Theorem}

\proof The proof of this theorem is very similar as the proof for the previous theorem. It follows the conclusion in Theorem~\ref{theorem_asymmetric1}.
\hfill\newQED

According to the theorem above, to calculate the upper bound of  nonuniform codes correcting $[t_{\downarrow}, t_{\uparrow}]$ errors,
we can first calculate the upper bound of nonuniform codes correcting $t_{\downarrow}+\underline{t_{\uparrow}}$ asymmetric errors. Generally speaking,
nonuniform codes correcting $[t_{\downarrow}, t_{\uparrow}]$ errors (considering the optimal case) are more efficient than nonuniform codes correcting $t_{\downarrow}+\overline{t_{\uparrow}}$ asymmetric errors,
but less efficient than those correcting $t_{\downarrow}+\underline{t_{\uparrow}}$ asymmetric errors.
According to the definitions of $\underline{t_{\uparrow}}$ and $\overline{t_{\uparrow}}(w)$, it is easy to get that
$$\underline{t_{\uparrow}}(w)\leq t_{\uparrow}(w) \leq \overline{t_{\uparrow}}(w),$$
for $0\leq w\leq n$.
Typically, if $p_{\downarrow},p_{\uparrow}\ll 1$, then $\overline{t_{\uparrow}}(w)-\underline{t_{\uparrow}}(w)\ll t_{\uparrow}(w)$.
It implies that
nonuniform codes correcting $[t_{\downarrow},t_{\uparrow}]$ errors are roughly as efficient as those correcting $t_{\downarrow}+t_{\uparrow}$ asymmetric errors.
If we consider i.i.d. errors and long codewords, it is equally difficult to correct
errors introduced by a binary asymmetric channel with crossover probabilities $p_{\downarrow}$ and $p_{\uparrow}$
or a Z-channel with a crossover probability $p_{\downarrow}+p_{\uparrow}$.

\section{Concluding Remarks}
\label{section_conclusion}

In storage systems with asymmetric errors, it is desirable to
design error-correcting codes such that the reliability of each codeword is
guaranteed in the worst case, and the size of the code is maximized.
This motivated us to propose the concept of nonuniform codes, whose codewords
can tolerate a number of asymmetric errors that depends on their Hamming weights.
We derived an almost explicit upper bound on the size of nonuniform codes and compared the
asymptotic performances of nonuniform codes and uniform codes - it is evident that there is
a potential performance gain by using nonuniform codes. In addition, we presented
two general constructions of nonuniform codes, including \emph{layered codes} and
\emph{flipping codes}. Open problems include efficient encoding for \emph{layered codes} and the
construction of \emph{flipping codes} when $p$ is not small. In general, the construction
of simple and efficient nonuniform codes is still an open problem.


%





\ifCLASSOPTIONcaptionsoff
  \newpage
\fi

\end{document}